\documentstyle[12pt,color,epsfig,ulem,subfigure]{article}
\def\baselinestretch{1.3}
\advance\textheight by \topskip
\newcommand{\comment}[1]{}

\def\beq{\begin{equation}}
\def\eeq{\end{equation}}
\def\beqn{\begin{eqnarray}}
\def\eeqn{\end{eqnarray}}

\def\mygraph#1#2{ \subfigure[]{
   \label{#1}
   \hspace*{-0.6in}
   \begin{minipage}[b]{0.5\textwidth}
   \centering
   \hspace*{4ex}
   \includegraphics[width=0.9\textwidth,height=0.7\textwidth]{#2}
   \vspace*{-6ex}
   \end{minipage}}
   \vspace*{-1ex}
}

\begin{document}
\tolerance=100000
\thispagestyle{empty}
\setcounter{page}{0}
\topmargin -0.1in
\headsep 30pt
\footskip 40pt
\oddsidemargin 12pt
\evensidemargin -16pt
\textheight 8.5in
\textwidth 6.5in
\parindent 20pt
 
\def\baselinestretch{1.5}
\newcommand{\newc}{\newcommand}
\def\preprint{{preprint}}
\def\Ord{\lower .7ex\hbox{$\;\stackrel{\textstyle <}{\sim}\;$}}
\def\OOrd{\lower .7ex\hbox{$\;\stackrel{\textstyle >}{\sim}\;$}}
\def\cO#1{{\cal{O}}\left(#1\right)}
\newc{\order}{{\cal O}}
\def\lag             {{\cal L}}
\def\Lag             {{\cal L}}
\def\lum             {{\cal L}}
\def\R               {{\cal R}}
\def\Rsq             {{\cal R}^{\sq}}
\def\Rst             {{\cal R}^{\st}}
\def\Rsb             {{\cal R}^{\sb}}
\def\M               {{\cal M}}
\def\Oas             {{\cal O}(\alpha_{s})}
\def\Vcal            {{\cal V}}
\def\Wcal            {{\cal W}}
\newc{\be}{\begin{equation}}
\newc{\ee}{\end{equation}}
\newc{\br}{\begin{eqnarray}}
\newc{\er}{\end{eqnarray}}
\newc{\ba}{\begin{array}}
\newc{\ea}{\end{array}}
\newc{\bi}{\begin{itemize}}
\newc{\ei}{\end{itemize}}
\newc{\bn}{\begin{enumerate}}
\newc{\en}{\end{enumerate}}
\newc{\bc}{\begin{center}}
\newc{\ec}{\end{center}}
\newc{\ul}{\underline}
\newc{\ol}{\overline}
\newc{\ra}{\rightarrow}
\newc{\lra}{\longrightarrow}
\newc{\wt}{\widetilde}
\newc{\til}{\tilde}
\def\kr              {^{\dagger}}
\newc{\wh}{\widehat}
\newc{\ti}{\times}
\newc{\Dir}{\kern -6.4pt\Big{/}}
\newc{\Dirin}{\kern -10.4pt\Big{/}\kern 4.4pt}
\newc{\DDir}{\kern -10.6pt\Big{/}}
\newc{\DGir}{\kern -6.0pt\Big{/}}
\newc{\sig}{\sigma}
\newc{\sigmalstop}{\sig_{\lstoppair}}
\newc{\Sig}{\Sigma}  
\newc{\del}{\delta}
\newc{\Del}{\Delta}
\newc{\lam}{\lambda}
\newc{\Lam}{\Lambda}
\newc{\gam}{\gamma}
\newc{\Gam}{\Gamma}
\newc{\eps}{\epsilon}
\newc{\Eps}{\Epsilon}
\newc{\kap}{\kappa}
\newc{\Kap}{\Kappa}
\newc{\modulus}[1]{\left| #1 \right|}
\newc{\eq}[1]{(\ref{eq:#1})}
\newc{\eqs}[2]{(\ref{eq:#1},\ref{eq:#2})}
\newc{\etal}{{\it et al.}\ }
\newc{\ibid}{{\it ibid}.}
\newc{\ibidem}{{\it ibidem}.}
\newc{\eg}{{\it e.g.}\ }
\newc{\ie}{{\it i.e.}\ }
\def \viz{\emph{viz.}}
\def \etc{\emph{etc. }}
\newc{\nonum}{\nonumber}
\newc{\lab}[1]{\label{eq:#1}}
\newc{\dpr}[2]{({#1}\cdot{#2})}
\newc{\lt}{\stackrel{<}}
\newc{\gt}{\stackrel{>}}
\newc{\lsimeq}{\stackrel{<}{\sim}}
\newc{\gsimeq}{\stackrel{>}{\sim}}
\def\lsim{\buildrel{\scriptscriptstyle <}\over{\scriptscriptstyle\sim}}
\def\gsim{\buildrel{\scriptscriptstyle >}\over{\scriptscriptstyle\sim}}
\def\lapp{\mathrel{\rlap{\raise.5ex\hbox{$<$}}
                    {\lower.5ex\hbox{$\sim$}}}}
\def\gapp{\mathrel{\rlap{\raise.5ex\hbox{$>$}}
                    {\lower.5ex\hbox{$\sim$}}}}
\newc{\half}{\frac{1}{2}}
\newcommand {\nnc}        {{\overline{\mathrm N}_{95}}}
\newcommand {\dm}         {\Delta m}
\newcommand {\dM}         {\Delta M}
\def\bra{\langle}
\def\ket{\rangle}
\def\cO#1{{\cal{O}}\left(#1\right)}
\def \DM{{\Delta{m}}}
\newc{\bQ}{\ol{Q}}
\newc{\dota}{\dot{\alpha }}
\newc{\dotb}{\dot{\beta }}
\newc{\dotd}{\dot{\delta }}
\newc{\nindnt}{\noindent}

\newcommand{\medf}[2] {{\footnotesize{\frac{#1}{#2}} }}
\newcommand{\smaf}[2] {{\textstyle \frac{#1}{#2} }}
\def\onesq            {{\textstyle \frac{1}{\sqrt{2}} }}
\def\onehf            {{\textstyle \frac{1}{2} }}
\def\oneth            {{\textstyle \frac{1}{3} }}
\def\twoth            {{\textstyle \frac{2}{3} }}
\def\onefo            {{\textstyle \frac{1}{4} }}
\def\forth            {{\textstyle \frac{4}{3} }}

\newc{\matth}{\mathsurround=0pt}
\def\ML{\ifmmode{{\mathaccent"7E M}_L}
             \else{${\mathaccent"7E M}_L$}\fi}
\def\MR{\ifmmode{{\mathaccent"7E M}_R}
             \else{${\mathaccent"7E M}_R$}\fi}
\newcommand{\s}{\\ \vspace*{-3mm} }

\def \ud { {1 \over 2} }
\def \ut { {1 \over 3} }
\def \td { {3 \over 2} }
\newc{\mr}{\mathrm}
\def\dh {\partial }
\def \cs { cross-section }
\def \css { cross-sections }
\def \cm { centre of mass }
\def \cms { centre of mass energy }
\def \cc { coupling constant }
\def \ccs {coupling constants }
\def \gc {gauge coupling }
\def \gcc {gauge coupling constant }
\def \gccs {gauge coupling constants }
\def \yc {Yukawa coupling }
\def \ycc {Yukawa coupling constant }
\def \pp {{parameter }}
\def \pps {{parameters }} 
\def \ps {parameter space }
\def \pss {parameter spaces }
\def \vv {vice versa }

\newc{\siminf}{\mbox{$_{\sim}$ {\small {\hspace{-1.em}{$<$}}}    }}
\newc{\simsup}{\mbox{$_{\sim}$ {\small {\hspace{-1.em}{$>$}}}    }}


\newc {\Zboson}{{\mathrm Z}^{0}}
\newc{\thetaw}{\theta_W}
\newc{\mbot}{{m_b}}
\newc{\mtop}{{m_t}}
\newc{\sm}{${\cal {SM}}$}
\newc{\as}{\alpha_s}
\newc{\aem}{\alpha_{em}}
\def \PI{{\pi^{\pm}}}
\newc{\ppbar}{\mbox{$p\ol{p}$}}
\newc{\bbbar}{\mbox{$b\ol{b}$}}
\newc{\ccbar}{\mbox{$c\ol{c}$}}
\newc{\ttbar}{\mbox{$t\ol{t}$}}
\newc{\eebar}{\mbox{$e\ol{e}$}}
\newc{\zzero}{\mbox{$Z^0$}}
\def \gamz{\Gam_Z}
\newc{\wplus}{\mbox{$W^+$}}
\newc{\wminus}{\mbox{$W^-$}}
\newc{\ellp}{\ell^+}
\newc{\ellm}{\ell^-}
\newc{\elp}{\mbox{$e^+$}}
\newc{\elm}{\mbox{$e^-$}}
\newc{\elpm}{\mbox{$e^{\pm}$}}
\newc{\qbar}     {\mbox{$\ol{q}$}}
\def \ewgroup{SU(2)_L \otimes U(1)_Y}
\def \smgroup{SU(3)_C \otimes SU(2)_L \otimes U(1)_Y}
\def \smcolorem{SU(3)_C \otimes U(1)_{em}}

\def \SSM  {Supersymmetric Standard Model}
\def \poincare{Poincare$\acute{e}$}
\def \superspace{\emph{superspace}}
\def \sfs{\emph{superfields}}
\def \superpot{\emph{superpotential}}
\def \csf{\emph{chiral superfield}}
\def \csfs{\emph{chiral superfields}}
\def \vsf{\emph{vector superfield }}
\def \vsfs{\emph{vector superfields}}
\newc{\Ebar}{{\bar E}}
\newc{\Dbar}{{\bar D}}
\newc{\Ubar}{{\bar U}}
\newc{\susy}{{{SUSY}}}
\newc{\msusy}{{{M_{SUSY}}}}

\def\photino{\ifmmode{\mathaccent"7E \gam}\else{$\mathaccent"7E \gam$}\fi}
\def\taugluino{\ifmmode{\tau_{\mathaccent"7E g}}
             \else{$\tau_{\mathaccent"7E g}$}\fi}
\def\mphotino{\ifmmode{m_{\mathaccent"7E \gam}}
             \else{$m_{\mathaccent"7E \gam}$}\fi}
\newc{\gl}   {\mbox{$\wt{g}$}}
\newc{\mgl}  {\mbox{$m_{\gl}$}}
\def \charginopm{{\wt\chi}^{\pm}}
\def \mcharginopm{m_{\charginopm}}
\def \mchpmmin {\mcharginopm^{min}}
\def \chonep {{\wt\chi_1^+}}
\def \ch2p {{\wt\chi_2^+}}
\def \chonem {{\wt\chi_1^-}}
\def \ch2m {{\wt\chi_2^-}}
\def \chplus {{\wt\chi^+}}
\def \chminus {{\wt\chi^-}}
\def \chonip{{\wt\chi_i}^{+}}
\def \chonim{{\wt\chi_i}^{-}}
\def \chonipm{{\wt\chi_i}^{\pm}}
\def \chonjp{{\wt\chi_j}^{+}}
\def \chonjm{{\wt\chi_j}^{-}}
\def \chonjpm{{\wt\chi_j}^{\pm}}
\def \chonepm{{\wt\chi_1}^{\pm}}
\def \chonemp{{\wt\chi_1}^{\mp}}
\def \mchonepm{m_{\chonepm}}
\def \mchonemp{m_{\chonemp}}
\def \chtwopm{{\wt\chi_2}^{\pm}}
\def \mchtwopm{m_{\chtwopm}}
\newc{\dmchi}{\Delta m_{\wt\chi}}


\def \vlsp{\emph{VLSP}}
\def \lspi{\wt\chi_i^0}
\def \mlspi{m_{\lspi}}
\def \lspj{\wt\chi_j^0}
\def \mlspj{m_{\lspj}}
\def \lspone{\wt\chi_1^0}
\def \mlspone{m_{\lspone}}
\def \lsptwo{\wt\chi_2^0}
\def \mlsptwo{m_{\lsptwo}}
\def \lspthree{\wt\chi_3^0}
\def \mlspthree{m_{\lspthree}}
\def \lspfour{\wt\chi_4^0}
\def \mlspfour{m_{\lspfour}}


\newc{\sele}{\wt{\mathrm e}}
\newc{\sell}{\wt{\ell}}
\def \msell{m_{\sell}}
\def \slepone{\wt\ell_1}
\def \mslepone{m_{\slepone}}
\def \smuone{\wt\mu_1}
\def \msmuone{m_{\smuone}}
\def \stauone{\wt\tau_1}
\def \mstauone{m_{\stauone}}
\def \snu{\wt{\nu}}
\def \snutau{\wt{\nu}_{\tau}}
\def \msnu{m_{\snu}}
\def \msnumu{m_{\snu_{\mu}}}
\def \barsnu{\wt{\bar{\nu}}}
\def \barsnul{\barsnu_{\ell}}
\def \snul{\snu_{\ell}}
\def \mbarsnu{m_{\barsnu}}
\newc{\snue}     {\mbox{$ \wt{\nu_e}$}}
\newc{\smu}{\wt{\mu}}
\newc{\stau}{\wt{\tau}}
\newc {\nuL} {\wt{\nu}_L}
\newc {\nuR} {\wt{\nu}_R}
\newc {\snub} {\bar{\wt{\nu}}}
\newc {\eL} {\wt{e}_L}
\newc {\eR} {\wt{e}_R}
\def \slepl{\wt{l}_L}
\def \mslepl{m_{\slepl}}
\def \slepr{\wt{l}_R}
\def \mslepr{m_{\slepr}}
\def \stau{\wt\tau}
\def \mstau{m_{\stau}}
\def \slepton{\wt\ell}
\def \mslepton{m_{\slepton}}
\def \mlhiggs{m_{h^0}}

\def \xr{X_{r}}

\def \sfer{\wt{f}}
\def \msfer{m_{\sfer}}
\def \sq{\wt{q}}
\def \msq{m_{\sq}}
\def \msquleft{m_{\tilde{u_L}}}
\def \msqurht{m_{\tilde{u_R}}}
\def \sql{\wt{q}_L}
\def \msql{m_{\sql}}
\def \sqr{\wt{q}_R}
\def \msqr{m_{\sqr}}
\newc{\msqot}  {\mbox{$m_(\sq_{1,2} )$}}
\newc{\sqbar}    {\mbox{$\bar{\wt{q}}$}}
\newc{\ssb}      {\mbox{$\squark\ol{\squark}$}}
\newc {\qL} {\wt{q}_L}
\newc {\qR} {\wt{q}_R}
\newc {\uL} {\wt{u}_L}
\newc {\uR} {\wt{u}_R}
\def \ul{\wt{u}_L}
\def \mul{m_{\ul}}
\newc {\dL} {\wt{d}_L}
\newc {\dR} {\wt{d}_R}
\newc {\cL} {\wt{c}_L}
\newc {\cR} {\wt{c}_R}
\newc {\sL} {\wt{s}_L}
\newc {\sR} {\wt{s}_R}
\newc {\tL} {\wt{t}_L}
\newc {\tR} {\wt{t}_R}
\newc {\stb} {\ol{\wt{t}}_1}
\newc {\sbot} {\wt{b}_1}
\newc {\msbot} {m_{\sbot}}
\newc {\sbotb} {\ol{\wt{b}}_1}
\newc {\bL} {\wt{b}_L}
\newc {\bR} {\wt{b}_R}
\def \mul{m_{\wt{u}_L}}
\def \mur{m_{\wt{u}_R}}
\def \mdl{m_{\wt{d}_L}}
\def \mdr{m_{\wt{d}_R}}
\def \mcl{m_{\wt{c}_L}}
\def \charml{\wt{c}_L}
\def \mcr{m_{\wt{c}_R}}
\newc{\csquark}  {\mbox{$\wt{c}$}}
\newc{\csquarkl} {\mbox{$\wt{c}_L$}}
\newc{\mcsl}     {\mbox{$m(\csquarkl)$}}
\def \msl{m_{\wt{s}_L}}
\def \msr{m_{\wt{s}_R}}
\def \mbl{m_{\wt{b}_L}}
\def \mbr{m_{\wt{b}_R}}
\def \mtl{m_{\wt{t}_L}}
\def \mtr{m_{\wt{t}_R}}
\def \st{\wt{t}}
\def \mst{m_{\st}}
\newc {\stopl}         {\wt{\mathrm{t}}_{\mathrm L}}
\newc {\stopr}         {\wt{\mathrm{t}}_{\mathrm R}}
\newc {\stoppair}      {\wt{\mathrm{t}}_{1}
\bar{\wt{\mathrm{t}}}_{1}}
\def \lstop{\wt{t}_{1}}
\def \lstopbar{\lstop^*}
\def \hstop{\wt{t}_{2}}
\def \hstopbar{\hstop^*}
\def \mlstop{m_{\lstop}}
\def \mhstop{m_{\hstop}}
\def \lstoppair{\lstop\lstop^*}
\def \hstoppair{\hstop\hstop^*}
\newc{\tsquark}  {\mbox{$\wt{t}$}}
\newc{\ttb}      {\mbox{$\tsquark\ol{\tsquark}$}}
\newc{\ttbone}   {\mbox{$\tsquark_1\ol{\tsquark}_1$}}
\def \tsq {top squark }
\def \tsqs {top squarks }
\def \tsql {ligtest top squark }
\def \tsqh {heaviest top squark }
\newc{\mix}{\theta_{\wt t}}
\newc{\cost}{\cos{\theta_{\wt t}}}
\newc{\sint}{\sin{\theta_{\wt t}}}
\newc{\costloop}{\cos{\theta_{\wt t_{loop}}}}
\def \lsbot{\wt{b}_{1}}
\def \lsbotbar{\lsbot^*}
\def \hsbot{\wt{b}_{2}}
\def \hsbotbar{\hsbot^*}
\def \mlsbot{m_{\lsbot}}
\def \mhsbot{m_{\hsbot}}
\def \lsbotpair{\lsbot\lsbot^*}
\def \hsbotpair{\hsbot\hsbot^*}
\newc{\mixsbot}{\theta_{\wt b}}

\def \mhone{m_{h_1}}
\def \hup{{H_u}}
\def \hdn{{H_d}}
\newc{\tb}{\tan\beta}
\newc{\cb}{\cot\beta}
\newc{\vev}[1]{{\left\langle #1\right\rangle}}

\def \abot{A_{b}}
\def \atop{A_{t}}
\def \atau{A_{\tau}}
\newc{\mhalf}{m_{1/2}}
\newc{\mzero} {\mbox{$m_0$}}
\newc{\azero} {\mbox{$A_0$}}

\newc{\lb}{\lam}
\newc{\lbp}{\lam^{\prime}}
\newc{\lbpp}{\lam^{\prime\prime}}
\newc{\rpv}{{\not \!\! R_p}}
\newc{\rpvm}{{\not  R_p}}
\newc{\rp}{R_{p}}
\newc{\rpmssm}{{RPC MSSM}}
\newc{\rpvmssm}{{RPV MSSM}}


\newc{\sbyb}{S/$\sqrt B$}
\newc{\pelp}{\mbox{$e^+$}}
\newc{\pelm}{\mbox{$e^-$}}
\newc{\pelpm}{\mbox{$e^{\pm}$}}
\newc{\epem}{\mbox{$e^+e^-$}}
\newc{\lplm}{\mbox{$\ell^+\ell^-$}}
\def \branch{\emph{BR}}
\def \branche{\branch(\lstop\ra be^{+}\nu_e \lspone)\ti \branch(\lstop^{*}\ra \bar{b}q\bar{q^{\prime}}\lspone)}
\def \branchmu{\branch(\lstop\ra b\mu^{+}\nu_{\mu} \lspone)\ti \branch(\lstop^{*}\ra \bar{b}q\bar{q^{\prime}}\lspone)}
\def\Ecm{\ifmmode{E_{\mathrm{cm}}}\else{$E_{\mathrm{cm}}$}\fi}
\newc{\rts}{\sqrt{s}}
\newc{\rtshat}{\sqrt{\hat s}}
\newc{\gev}{\,GeV}
\newc{\mev}{~{\rm MeV}}
\newc{\tev}  {\mbox{$\;{\rm TeV}$}}
\newc{\gevc} {\mbox{$\;{\rm GeV}/c$}}
\newc{\gevcc}{\mbox{$\;{\rm GeV}/c^2$}}
\newc{\intlum}{\mbox{${ \int {\cal L} \; dt}$}}
\newc{\call}{{\cal L}}
\def \met  {\mbox{${E\!\!\!\!/_T}$}}
\def \cpv  {\mbox{${CP\!\!\!\!/}$}}
\newc{\ptmiss}{/ \hskip-7pt p_T}
\def \eslash{\not \! E}
\def \etslash{\not \! E_T }
\def \ptslash{\not \! p_T }
\newc{\PT}{\mbox{$p_T$}}
\newc{\ET}{\mbox{$E_T$}}
\newc{\dedx}{\mbox{${\rm d}E/{\rm d}x$}}
\newc{\ifb}{\mbox{${\rm fb}^{-1}$}}
\newc{\ipb}{\mbox{${\rm pb}^{-1}$}}
\newc{\pb}{~{\rm pb}}
\newc{\fb}{~{\rm fb}}
\newc{\ycut}{y_{\mathrm{cut}}}
\newc{\chis}{\mbox{$\chi^{2}$}}
\def \hadron{\emph{hadron}}
\def \nlc{\emph{NLC }}
\def \lhc{\emph{LHC }}
\def \cdf{\emph{CDF }}
\def\dzero{\emptyset}
\def \tevatron{\emph{Tevatron }}
\def \lep{\emph{LEP }}
\def \jets{\emph{jets }}
\def \jet(s){\emph{jet(s) }}

\def\Crs{stroke [] 0 setdash exch hpt sub exch vpt add hpt2 vpt2 neg V currentpoint stroke 
hpt2 neg 0 R hpt2 vpt2 V stroke}
\def\loopdk{\lstop \ra c \lspone}
\def\brloopdk{\branch(\loopdk)}
\def\fourdk{\lstop \ra b \lspone  f \bar f'}
\def\brfourdk{\branch(\fourdk)}
\def\fourdklep{\lstop \ra b \lspone  \ell \nu_{\ell}}
\def\fourdkhad{\lstop \ra b \lspone  q \bar q'}
\def\brfourdklep{\branch(\fourdklep)}
\def\brfourdkhad{\branch(\fourdkhad)}
\def\twodk{\lstop \ra b \chonep}
\def\brtwodk{\branch(\twodk)}
\def\threedkslep{\lstop \ra b \wt{\ell} \nu_{\ell}}
\def\brthreedkslep{\branch(\threedkslep)}
\def\threedksnu{\lstop \ra b \wt{\nu_{\ell}} \ell }
\def\brthreedksnu{\branch(\threedksnu) }
\def\threedklsp{\lstop \ra b W \lspone }
\def\brthreedklsp{\\branch(\threedklsp) }
\def\topdk{t \ra \lstop \lspone}
\def\rpvdk{\lstop \ra e^+ d}
\def\brrpvdk{\branch(\rpvdk)}
\def\fonec{f_{11c}} 
\newc{\mpl}{M_{\rm Pl}}
\newc{\mgut}{M_{GUT}}
\newc{\mw}{M_{W}}
\newc{\mweak}{M_{weak}}
\newc{\mz}{M_{Z}}

\newc{\OPALColl}   {OPAL Collaboration}
\newc{\ALEPHColl}  {ALEPH Collaboration}
\newc{\DELPHIColl} {DELPHI Collaboration}
\newc{\XLColl}     {L3 Collaboration}
\newc{\JADEColl}   {JADE Collaboration}
\newc{\CDFColl}    {CDF Collaboration}
\newc{\DXColl}     {D0 Collaboration}
\newc{\HXColl}     {H1 Collaboration}
\newc{\ZEUSColl}   {ZEUS Collaboration}
\newc{\LEPColl}    {LEP Collaboration}
\newc{\ATLASColl}  {ATLAS Collaboration}
\newc{\CMSColl}    {CMS Collaboration}
\newc{\UAColl}    {UA Collaboration}
\newc{\KAMLANDColl}{KamLAND Collaboration}
\newc{\IMBColl}    {IMB Collaboration}
\newc{\KAMIOColl}  {Kamiokande Collaboration}
\newc{\SKAMIOColl} {Super-Kamiokande Collaboration}
\newc{\SUDANTColl} {Soudan-2 Collaboration}
\newc{\MACROColl}  {MACRO Collaboration}
\newc{\GALLEXColl} {GALLEX Collaboration}
\newc{\GNOColl}    {GNO Collaboration}
\newc{\SAGEColl}  {SAGE Collaboration}
\newc{\SNOColl}  {SNO Collaboration}
\newc{\CHOOZColl}  {CHOOZ Collaboration}
\newc{\PDGColl}  {Particle Data Group Collaboration}

\def\issue(#1,#2,#3){{\bf #1}, #2 (#3)}
\def\ASTR(#1,#2,#3){Astropart.\ Phys. \issue(#1,#2,#3)}
\def\AJ(#1,#2,#3){Astrophysical.\ Jour. \issue(#1,#2,#3)}
\def\AJS(#1,#2,#3){Astrophys.\ J.\ Suppl. \issue(#1,#2,#3)}
\def\APP(#1,#2,#3){Acta.\ Phys.\ Pol. \issue(#1,#2,#3)}
\def\JCAP(#1,#2,#3){Journal\ XX. \issue(#1,#2,#3)} 
\def\SC(#1,#2,#3){Science \issue(#1,#2,#3)}
\def\PRD(#1,#2,#3){Phys.\ Rev.\ D \issue(#1,#2,#3)}
\def\PR(#1,#2,#3){Phys.\ Rev.\ \issue(#1,#2,#3)} 
\def\PRC(#1,#2,#3){Phys.\ Rev.\ C \issue(#1,#2,#3)}
\def\NPB(#1,#2,#3){Nucl.\ Phys.\ B \issue(#1,#2,#3)}
\def\NPPS(#1,#2,#3){Nucl.\ Phys.\ Proc. \ Suppl \issue(#1,#2,#3)}
\def\NJP(#1,#2,#3){New.\ J.\ Phys. \issue(#1,#2,#3)}
\def\JP(#1,#2,#3){J.\ Phys.\issue(#1,#2,#3)}
\def\PL(#1,#2,#3){Phys.\ Lett. \issue(#1,#2,#3)}
\def\PLB(#1,#2,#3){Phys.\ Lett.\ B  \issue(#1,#2,#3)}
\def\ZP(#1,#2,#3){Z.\ Phys. \issue(#1,#2,#3)}
\def\ZPC(#1,#2,#3){Z.\ Phys.\ C  \issue(#1,#2,#3)}
\def\PREP(#1,#2,#3){Phys.\ Rep. \issue(#1,#2,#3)}
\def\PRL(#1,#2,#3){Phys.\ Rev.\ Lett. \issue(#1,#2,#3)}
\def\MPL(#1,#2,#3){Mod.\ Phys.\ Lett. \issue(#1,#2,#3)}
\def\RMP(#1,#2,#3){Rev.\ Mod.\ Phys. \issue(#1,#2,#3)}
\def\SJNP(#1,#2,#3){Sov.\ J.\ Nucl.\ Phys. \issue(#1,#2,#3)}
\def\CPC(#1,#2,#3){Comp.\ Phys.\ Comm. \issue(#1,#2,#3)}
\def\IJMPA(#1,#2,#3){Int.\ J.\ Mod. \ Phys.\ A \issue(#1,#2,#3)}
\def\MPLA(#1,#2,#3){Mod.\ Phys.\ Lett.\ A \issue(#1,#2,#3)}
\def\PTP(#1,#2,#3){Prog.\ Theor.\ Phys. \issue(#1,#2,#3)}
\def\RMP(#1,#2,#3){Rev.\ Mod.\ Phys. \issue(#1,#2,#3)}
\def\NIMA(#1,#2,#3){Nucl.\ Instrum.\ Methods \ A \issue(#1,#2,#3)}
\def\JHEP(#1,#2,#3){J.\ High\ Energy\ Phys. \issue(#1,#2,#3)}
\def\EPJC(#1,#2,#3){Eur.\ Phys.\ J.\ C \issue(#1,#2,#3)}
\def\RPP (#1,#2,#3){Rept.\ Prog.\ Phys. \issue(#1,#2,#3)}
\def\PPNP(#1,#2,#3){ Prog.\ Part.\ Nucl.\ Phys. \issue(#1,#2,#3)}
\newc{\PRDR}[3]{{Phys. Rev. D} {\bf #1}, Rapid  Communications, #2 (#3)}

\vspace*{\fill}
\vspace{-0.8in}
\begin{flushright}
\end{flushright}
\begin{center}
{\Large \bf
Non-zero trilinear parameter in the mSUGRA model - dark matter
and collider signals at Tevatron and 
LHC}
  \vglue 0.5cm
  Utpal Chattopadhyay$^{(a)}$\footnote{tpuc@iacs.res.in}, 
Debottam Das$^{(a)}$\footnote{tpdd@iacs.res.in},
Amitava Datta$^{(b)}$\footnote{adatta@juphys.ernet.in} and 
Sujoy Poddar$^{(b)}$\footnote{sujoy@juphys.ernet.in}
    \vglue 0.2cm
    {\it $^{(a)}$Department of Theoretical Physics, Indian Association
for the Cultivation of Science, Raja S.C. Mullick Road, Kolkata 700 032, India \\}
    {\it $^{(b)}$
Department of Physics, Jadavpur University, Jadavpur, Kolkata 700 032, India
\\}
\end{center}

\begin{abstract}
{\noindent \normalsize}
Phenomenologically viable and 
interesting regions of parameter space in the minimal super-gravity (mSUGRA) 
model with small $m_0$ and small $m_{1/2}$ consistent with the WMAP data 
on dark 
matter relic density 
and the bound on the mass of the lightest Higgs scalar $ m_h>$ 114 GeV 
from LEP2
open up if the rather adhoc assumption $A_0$=0, where 
$A_0$ is the common trilinear soft breaking parameter, employed in most of 
the existing analyses is relaxed.  Since this region corresponds to 
relatively light squarks and gluinos which are likely to be probed extensively in 
the very early stages of the LHC experiments, the consequences of moderate 
or large negative values of $A_0$ are examined in detail.  We find that in 
this region several processes including lightest supersymmetric particle 
(LSP) pair annihilation, LSP - lighter tau slepton (${\tilde \tau}_1$) 
coannihilation and LSP - lighter top squark (${\tilde t}_1$) 
coannihilation contribute to the observed dark matter relic density.  
The possibility that a ${\tilde t}_1$ that can participate in 
coannihilation 
with the lightest neutralino to satisfy the WMAP bound on relic density 
and at the same time  be 
observed at the current experiments at the Tevatron is wide open.
 At the 
LHC a large number of squark - gluino events lead to a very distinctive 
semi-inclusive signature $\tau^\pm$+X$_\tau$ (anything without a tau 
lepton) with a characteristic size much larger than $e^\pm$+X$_e$ or 
$\mu^\pm$+X$_\mu$ events. 
\end{abstract}

\newpage

\section{Introduction} \label{intro4} 

Models with supersymmetry(SUSY)\cite{SUSY} are interesting for a variety of
reasons.  Theoretically the removal of quadratic divergence in the Higgs boson
mass in the standard model (SM) by similar divergent loop diagrams involving
SUSY particles (sparticles) is very attractive.  These models have also
predicted very interesting experimental signatures and are very promising
candidates for beyond the standard model physics.  A specially attractive
feature of the minimal supersymmetric standard model (MSSM) with R-parity
conservation is the presence of the stable, weakly interacting lightest
neutralino (${\tilde \chi}_1^0$ )~\cite{goldbergDM} which turns out to be a
very good candidate for the observed cold dark matter (CDM) in the universe.
\cite{DMreviewgeneral,DMreview,recentSUSYDMreview}.

Since superpartners are yet to be observed supersymmetry must be a broken 
symmetry and 
one requires a soft SUSY breaking mechanism that preserves 
gauge invariance and does not give rise to any quadratic divergence. 
The MSSM has a large number of such soft breaking 
parameters. 
Being guided by well motivated theoretical ideas as well as 
low energy phenomenology, models with specified SUSY breaking 
mechanisms drastically reduces the large number of such arbitrary 
parameters to only a few. 
The $N=1$ supergravity models incorporate gravity mediated supersymmetry 
breaking and the models are attractive because of many features like gauge coupling unification, 
radiative breaking of electroweak symmetry \cite{rewsbrefs}, controlling flavour changing neutral 
currents (FCNC) by specific and simple assumptions at the unification 
scale \cite{SUSY}.

The simplest gravity mediated SUSY breaking model - the minimal supergravity 
(mSUGRA)\cite{msugra} model has only five free parameters. These are the common gaugino and scalar 
mass parameters $m_{1/2}$ and $m_0$, the common tri-linear coupling 
parameter $A_0$, all given at the gauge coupling unification scale 
($M_G \sim 2 \times 10^{16}$~GeV), the ratio of Higgs vacuum expectation 
values at the electroweak scale namely $\tan\beta$ which is in fact largely independent of scale 
and the sign of $\mu$, the higgsino mixing 
parameter. The magnitude of $\mu$ is obtained by the radiative 
electroweak symmetry breaking (REWSB) mechanism\cite{rewsbrefs}.
  The low energy sparticle spectra and couplings at the electroweak scale 
are generated by renormalization group evolutions (RGE) of soft 
breaking masses and coupling 
parameters\cite{oldrge,newrge}.    

Various SUSY models have been constrained by the data 
on cold dark matter relic density. But the
recent resurgence of interest in this field is due to the very restrictive
data from the Wilkinson Microwave Anisotropy Probe (WMAP) observation\cite{WMAPdata}. Combining the WMAP data with the results from the SDSS (Sloan Digital
Sky Survey) one obtains the conservative 3 $\sigma$ limits 

\begin{equation}
0.09 < \Omega_{CDM}h^2 < 0.13
\label{relicdensity}
\end{equation}

\noindent
where $\Omega_{CDM}h^2$ is is the DM relic density in units of the critical 
density, $h=0.71\pm0.026$ is the Hubble constant in units of
$100 \ \rm Km \ \rm s^{-1}\ \rm Mpc^{-1}$.
In supergravity type of models ${\tilde \chi}_1^0$
becomes the LSP for most of the 
parameter space\cite{DMreview,recentSUSYDMreview} 
and one may consider
$\Omega_{CDM}\equiv 
\Omega_{{\tilde \chi}_1^0}$. 
We should note here that the upper bound of
$\Omega_{{\tilde \chi}_1^0}$ in Eq.(\ref{relicdensity}) is a strong limit but lower
bound becomes weaker if we accept other candidates of dark matter.

In the thermally generated dark matter scenario, at very high 
temperature of the early universe
($T>>m_{{\tilde \chi}_1^0}$), ${\tilde \chi}_1^0$ 
was in thermal equilibrium with its annihilation products.  
The annihilation products include fermion pairs ($f \bar f$), 
gauge boson pairs ($W^+W^-$ \& $ZZ$), Higgs boson pairs ($hh,HH,AA,hH,hA,HA,H^+H^-$) or 
gauge boson-Higgs boson pairs
($Zh,ZH,ZA$~\&~$W^\pm W^\mp$) through $s,t$ and $u$ channel
diagrams.  Thereafter, at lower temperatures the annihilation rate
falls below the expansion rate of the universe and ${\tilde \chi}_1^0$
goes away from thermal equilibrium (freeze-out).
The present value of the $\Omega_{{\tilde \chi}_1^0}h^2$ can thus be computed
by solving the Boltzmann equation for $n_{{\tilde \chi}_1^0}$, the number 
density of the LSP in a Friedmann-Robertson-Walker universe.  Finding the 
neutralino
relic density most importantly involves computing the thermally 
averaged quantity $<\sigma_{eff} v>$, where 
$v$ is the relative velocity between two neutralinos annihilating 
each other and 
$\sigma_{eff}$ is the neutralino annihilation cross section which 
involves all possible final states. In addition to the annihilation of a  
LSP pair, coannihilations of the LSP 
\cite{coannistau,coannistop,Edsjo:1997bg,mizuta,
coanniSet2} may also be important. This happens 
if there are sparticles with masses approximately degenerate with the LSP mass. 

The annihilation cross section $\sigma_{eff}$ 
depends on the magnitude of the bino ($\tilde B$),the wino ($\tilde W$) and 
the Higgsino (${\tilde H}_1^0$ , ${\tilde H}_2^0$) compositions of the 
lightest neutralino and proximity in mass of the LSP with any coannihilating 
sparticle. Generically the
LSP composition is given by the following mixing: 
\begin{equation}
{\tilde \chi}_1^0=N_{11}\tilde B + N_{12}{\tilde W}_3 +
N_{13}{\tilde H}_1^0 +N_{14}{\tilde H}_2^0. 
\end{equation}
Here the coefficients $N_{ij}$  are the elements of 
the matrix that diagonalizes the neutralino mass matrix.  
One typically quantifies the 
composition through the 
gaugino fraction of ${\tilde \chi}_1^0$ which is defined as $F_g=|N_{11}|^2 +
|N_{12}|^2$. A ${\tilde \chi}_1^0$ would be called gaugino like if
$F_g$ is very close to 1($\gsim 0.9$) , higgsino like if $F_g \lsim 0.1$.
Otherwise the LSP would be identified as a gaugino-higgsino mixed state.

Accordingly a typical MSSM parameter space 
(with gaugino mass universality) 
where the WMAP constraint is satisfied can be
classified into several regions depending on the 
LSP annihilation/coannihilation mechanism. The list goes as follows. 

 i) In the so called {\it stau coannihilation} region, a 
large degree of ${\tilde \chi}_1^0-{\tilde \tau}_1$ 
coannihilation\cite{coannistau}  
reduces the thermal abundance  
sufficiently so as to satisfy the 
WMAP data.  The WMAP allowed region in the $m_0-m_{1/2}$ plane in the mSUGRA model typically 
falls near the boundary of the forbidden parameter space where ${\tilde \tau}_1$ becomes the LSP.  
The coannihilation processes are of the type 
${\tilde \chi}_1^0 \tilde \ell_R^a
\rightarrow \ell^a \gamma, \ell^a Z, \ell^a h$,
$\tilde \ell_R^a \tilde \ell_R^b \rightarrow \ell^a \ell^b$,
 and $\tilde \ell_R^a \tilde \ell_R^{b*} \rightarrow \ell^a\bar \ell^b,
\gamma \gamma, \gamma Z, ZZ, W^+W^-, hh$. Here $\tilde l$ is 
essentially the ${\tilde \tau}_1$.

ii) The {\it focus point}\cite{focus} or the {\it Hyperbolic branch}\cite{hyper} 
region in the mSUGRA model is characterized by a reduced value of $|\mu|$.  A small $|\mu|$ 
causes the LSP to have a significant Higgsino component or it can even 
be a pure Higgsino.
Strong coannihilation of the LSP with lighter chargino 
${\tilde \chi}_1^\pm$ occurs in this zone.  
Some of the dominant coannihilation 
processes in these region are\cite{Edsjo:1997bg,mizuta}:
${\tilde \chi}_1^0 {\tilde \chi}_1^{+}, {\tilde \chi}_2^0
{\tilde \chi}_1^{+}\rightarrow u_i\bar d_i, \bar e_i\nu_i, AW^+,ZW^+,
W^+h$; ${\tilde \chi}_1^{+} {\tilde \chi}_1^{-}, {\tilde \chi}_1^0 {\tilde 
\chi}_2^{0}\rightarrow u_i\bar u_i, d_i \bar d_i,
W^+W^-$.  Having the smallest mass difference between 
coannihilating sparticles
the process ${\tilde \chi}_1^0 {\tilde \chi}_1^{+}$ indeed dominates among
the above channels.
As a result the thermal abundance 
is reduced appreciably so that it satisfies
the WMAP data or coannihilations may even reduce it further (below the lower
limit of Eq.\ref{relicdensity}) thus causing ${\tilde \chi}_1^0$
to be a sub dominant component of dark matter.

iii) The {\it funnel or the Higgs-pole} region\cite{dreesDM93,funnel} 
satisfies the WMAP data 
for large values of $\tan\beta$ extending to both large $m_0$ 
and large $m_{1/2}$ regions. This is characterized by the direct-channel pole 
$2m_{{\tilde \chi}_1^0} \sim m_A,m_H$.  

iv)  For a limited range of large negative values of $A_0$, 
the lighter stop can become 
very light so that it may coannihilate with the LSP. This is the 
{\it stop coannihilation} region\cite{coannistop} 
characterized by a very light $m_{{\tilde t}_1}$.

v) In the {\it bulk annihilation region} or the bulk region\cite{coannistau}
where $m_0$ and $m_{1/2}$ are reasonably small in mSUGRA,
the LSP turns out to be bino dominated and,
consequently, couples favourably to right sleptons, which in fact
are the lightest sfermions in this region of parameter space.
As a result an LSP pair annihilates into SM fermions via  the 
exchange of light sfermions 
in the t-channel. This cross section depends on the mass of the LSP
($\mlspone$), its coupling with the sfermions and the masses of the
exchanged sfermions \cite{dreesDM93,DMreview,recentSUSYDMreview}.
However, LEP2 has imposed strong bounds on sparticle masses\cite{limits}, 
particularly 
on the slepton masses in the present context of mSUGRA and 
typical studies with $A_0=0$ disfavors a part of the bulk annihilation zone. 
Additionally, a very severe restriction
 appears on the $(m_0-m_{1/2})$ plane of mSUGRA 
from the bound on lightest Higgs boson mass ($m_h$) 
\cite{hlim} which  practically 
rules out the entire annihilation region. 
For $m_h >$114 GeV one finds that within the frame work of mSUGRA 
the sleptons are significantly  
heavier than the direct LEP2 bounds  and this leads to very small LSP 
annihilation
cross section implying an unacceptably large relic density. 
Thus it has often been claimed in the 
recent literature that the mSUGRA parameter space with low values of both 
$m_0$ and $\mhalf$ is strongly disfavoured. This automatically eliminates
the bulk annihilation region \cite{dmsugra}.

For small $m_0$ but relatively large $\mhalf$ one obtains a parameter space 
consistent with the Higgs mass bound. In this region a large $\mhalf$ pushes 
up the mass of the lightest neutralino while the lighter tau slepton 
($\stauone$) mass 
is rather modest because of a small $m_0$. Thus the LSP and $\stauone$ are
approximately mass degenerate and LSP-$\stauone$ coannihilation
\cite{coannistau} 
provides a viable mechanism of producing an acceptable relic density. 
However, the relatively large $\mhalf$ tends to push
up the masses of squarks and gluino and this reduces the size of the LHC 
signals significantly. 

The purpose of this paper is to emphasize that the above conclusions are
artifacts of the adhoc choice $A_0 = 0$. Many of the current analyses invoke
this choice without any compelling theoretical or experimental reason. On the
other hand it is well known that moderate to large negative\footnote{
We follow the standard sign convention of Ref.\cite{signconvention} for the
signs of $\mu$ and $A_0$.} values of $A_0$
lead to larger $m_h \cite{carenaEtc}$. Hence in this case the bound on $m_h$ can be satisfied
even for relatively small $m_0$ and small $\mhalf$. This revives the region where
LSP pair annihilation is significant, which would otherwise remain forbidden
for $A_0 = 0$ ( see, e.g., LEPSUSY working group figures in
Ref.\cite{lepsusy}) 
\footnote{For $A_0>0$, one requires $m_0$ and $\mhalf$ typically larger 
than the corresponding values for $A_0$=0. This does not lead to any novel 
collider signal.}.
Moreover the low $m_0-m_{1/2}$ regions of the mSUGRA
parameter space are characterized by relatively light squarks and gluino.  
Thus this region will be extensively probed at the early stages of the LHC
experiments.  It is therefore worthwhile to analyze the anticipated collider
signals corresponding to this region in detail.

Even for moderate negative values of $A_0$ the WMAP allowed 
parameter space extends 
considerably.  More importantly 
a small but interesting region where LSP pair annihilation 
produces an acceptable 
dark matter relic density
is revived. There is also a region where LSP - $\stauone$ 
coannihilation is still the most important
mechanism for having observed dark matter of the universe. Remarkably, even this region 
corresponds to a much smaller $\mhalf$ compared to what one would obtain for 
the $A_0$ =0 case. 
Consequently the squarks and gluinos become  relatively lighter.  
Furthermore large negative values of $A_0$ leads to a 
relatively light top squark with obvious characteristics in the collider 
signals. 
In fact for a small region of the parameter space, the LSP- 
$\lstop$ coannihilation\cite{coannistop} may significantly 
contribute to the observed dark matter density.

In this paper 
emphasis will be given on the features
of the sparticle spectrum and signals at the Tevatron and the LHC
corresponding to
the WMAP allowed  regions of the parameter space 
opened up by non-zero $A_0$.  Such signals will be
compared and contrasted with the expectations from the scenarios with $A_0 
= 0$. 

Several earlier dark matter analyses considered non-zero trilinear
couplings in various SUSY models including mSUGRA\cite{severalnonzeroa0}. 
More recently it has also been noted in the literature
\cite{newbulk} that the DM allowed mSUGRA parameter space is expanded for 
non-zero $A_0$ inspite of the constraint in Eq.1 and the lower bound
 on $m_h$ from LEP2. 
However, in this work we go further and point out the dominant 
annihilation/co-annihilation mechanisms that would 
produce the acceptable amount of neutralino  
relic density in different regions 
and analyze the novel collider signals associated with them.

We have mainly considered the direct constraints from LEP2 searches and WMAP 
data
on the mSUGRA parameter space. Other theoretical and experimental indirect
constraints have also been considered in the literature. For example, it
is well known that large values of $A_0$ may lead to a charge and colour
breaking (CCB) minima\cite{casasCCB}.  We shall comment on it in the next section. 
  
We note in passing that the constraints from the observed branching ratio 
BR($b -> s \gamma$)\cite{bsg-recent} 
disfavours the mSUGRA parameter space where both $m_0$ and
$\mhalf$ happen to be small\cite{dmsugra}. Parts of 
parameter space at the focus 
of
attention of this paper belongs to this category.  However, the
theoretical prediction of this branching ratio in mSUGRA is based on the 
assumption of
perfect alignment of the squark and quark mass matrices. This essentially
means that the mixing angle factors at certain vertices ( e.g., the
quark-squark- gaugino vertices) are identical to the Cabibbo-Kobayashi-Maskawa 
(CKM) factors at the
corresponding SM vartices. This assumption is unnecessarily restrictive
and may be falsified  by small off diagonal terms in the squarks mass 
matrices
at $M_G$ which may change the mixing pattern in the squark sector drastically. 
It should be emphasized that the above small off diagonal
entries do not affect the flavour conserving processes like neutralino annihilation or
events at hadron colliders in any significant way.
For a brief review of the model
dependent assumptions in the $b -> s \gamma$ analyses we refer the reader 
to Okumura et al in \cite{bsg-okumura} and Djouadi et al in \cite{dmsugra}.
 

\section{The DM Allowed Parameter Space for Non-zero $A_0$ and the 
Sparticle Spectrum}
\label{numerical}

In this section we relax the rather arbitrary 
choice $A_0=0$ and 
reexamine the scenarios with  small $m_0$ and small $m_{1/2}$ consistent 
with the WMAP data. We remind ourselves that this particular region 
is strongly disfavored by the $m_h$ bound from LEP2 for 
vanishing $A_0$. 

We list the non-trivial consequences of considering a non-vanishing $A_0$.  First,
the lightest Higgs boson becomes heavier even if we choose a moderately negative
value for $A_0$ keeping all other mSUGRA input parameters fixed.  This happens via
the mixing effects in the radiative corrections to $m_h$ \cite{carenaEtc}.  
Moderately large negative values like $A_0\approx -500$ GeV are adequate to satisfy the experimental lower bound on $m_h$ even for small $m_0$ and small
$m_{1/2}$.

However, three important additional effects should be taken into account while
exploring the consequences of non-vanishing $A_0(<0)$. i) With a negative $A_0$
having moderately large magnitude (thereby $A_t$ - the trilinear coupling at the
top quark sector at the weak scale being further negative), the lighter
top squark ${\tilde t}_1$
will become much lighter than the other squarks because of the off-diagonal
element $m_t(A_t-\mu\cot\beta)$ in the top squark mass matrix.  In fact 
${\tilde t_1}$ may even become the next lightest super particle (NLSP) or 
even violate the
current experimental lower limit if $A_t$ is sufficiently large and negative. 
We note here
that unless $\tan\beta$ is very small, the amount of mixing in the ${\tilde t}_1$
sector is dominated by $A_t$ in spite of the fact that $|\mu|$ determined by REWSB
also becomes larger as $A_0$ is driven towards large negative values.  ii) In the
${\tilde \tau}$ sector ${\tilde \tau}_1$ becomes lighter for large 
$|A_0|$ for
entirely different reason.  Here $\mu\tan\beta$ dominates over $A_\tau$ even for
large $|A_0|$ and moderate tan $\beta$.  This pushes ${\tilde \tau}_1$ to 
become lighter.  The effect of mixing is obviously larger for larger $\tan\beta$ or a
larger $m_{1/2}$. As a result ${\tilde \tau}_1$ and to a lesser extent ${\tilde
t}_1$ may become NLSP in different zones of the parameter space around the bulk
annihilation region. These two sparticles may thus coannihilate with the LSP and
consistency with the WMAP limits on relic density may be obtained. iii) As
mentioned in the introduction when $|A_0|$ becomes large one may hit a CCB minima
of the scalar potential\cite{casasCCB}.  We have imposed the CCB conditions at the 
electroweak
scale in this analysis.  However these constraints can be altogether relaxed if
it is assumed that the universe is built on the EWSB false vacuum with a life time
larger than the age of the universe \cite{false}.  We therefore do not pursue 
these constraints at other scales.


In this analysis we have generated the electroweak scale sparticle 
spectrum from the input parameters at the unification scale 
by using the SuSpect code\cite{suspect} with including the CCB condition as
mentioned before. This code employs the
two-loop renormalization group equations and 
implements the radiative electroweak symmetry 
breaking mechanism.  We have used the code micrOMEGAs\cite{micromegas} for 
computing the neutralino relic density. 
We have included LEP2 lower limits for sparticle masses\cite{limits} and 
set top mass $m_t =$173 GeV. 
In particular the lower bound  94~GeV on $m_{\stauone}$ is quite important.  
However, we should also
note here that the Higgs 
mass bound $m_h >$ 114 GeV should be treated with 
caution. The  theoretical prediction for the Higgs mass in the MSSM has been 
computed including two loop corrections\cite{carenaEtc}.  
It is well known that this  prediction  
involves an uncertainty of about 3~GeV \cite{sven1,sven2,sven3,sven4,djouadi}.
The sources of this uncertainty are the
momentum-independent two-loop corrections, the
momentum-dependent two-loop corrections, the higher loop corrections
from t/$\wt t $ sector etc. While 
presenting the allowed parameter space we have , therefore, delineated the 
regions corresponding to 111~GeV $<m_h<$ 114~GeV. 
This parameter space which cannot be presently ruled out with certainty will be 
referred to as the
uncertain $m_h$ zone in this paper.

\begin{figure}[!ht]
\vspace*{-0.1in}
\mygraph{t10m0120}{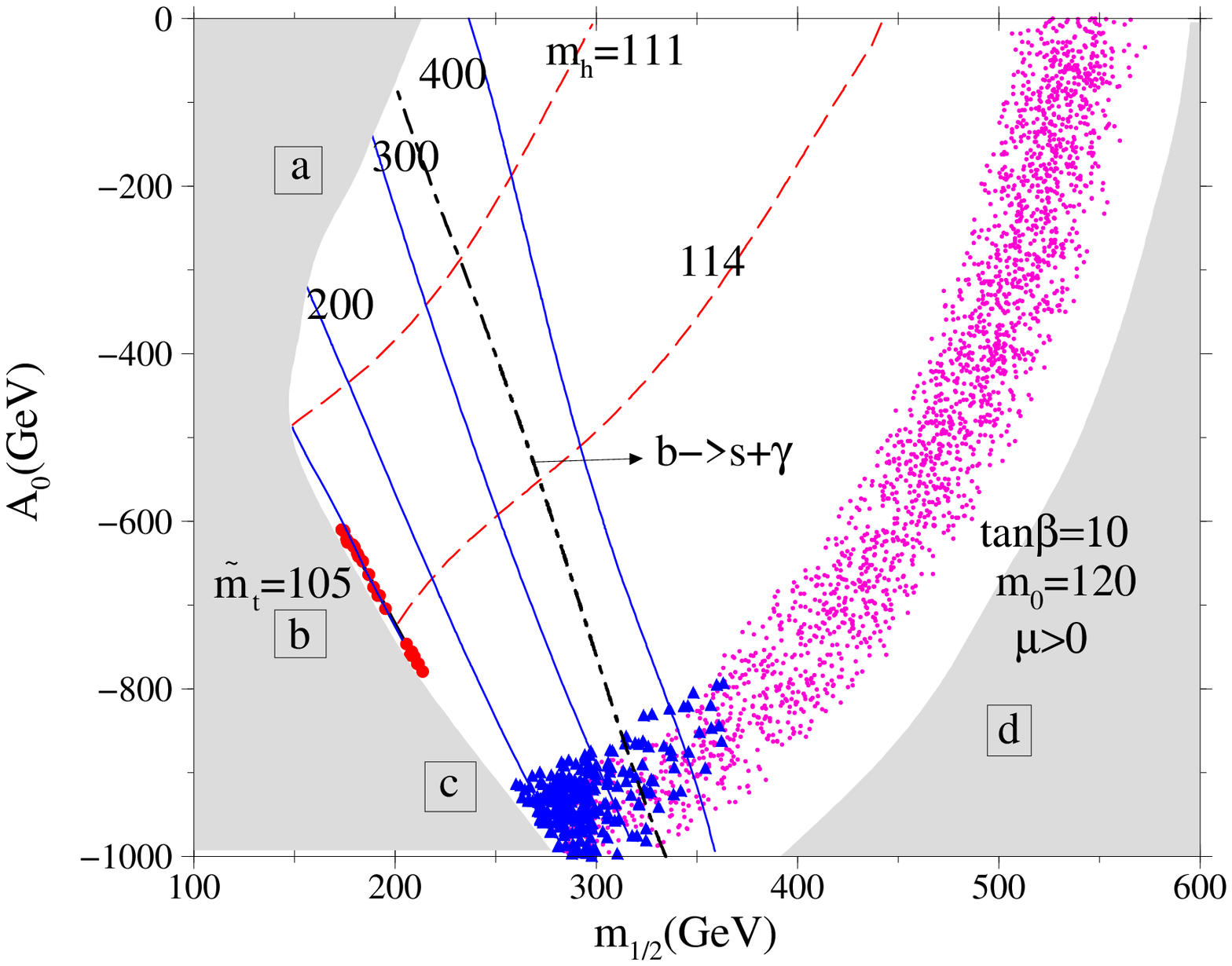}
\hspace*{0.5in}
\mygraph{t10m080}{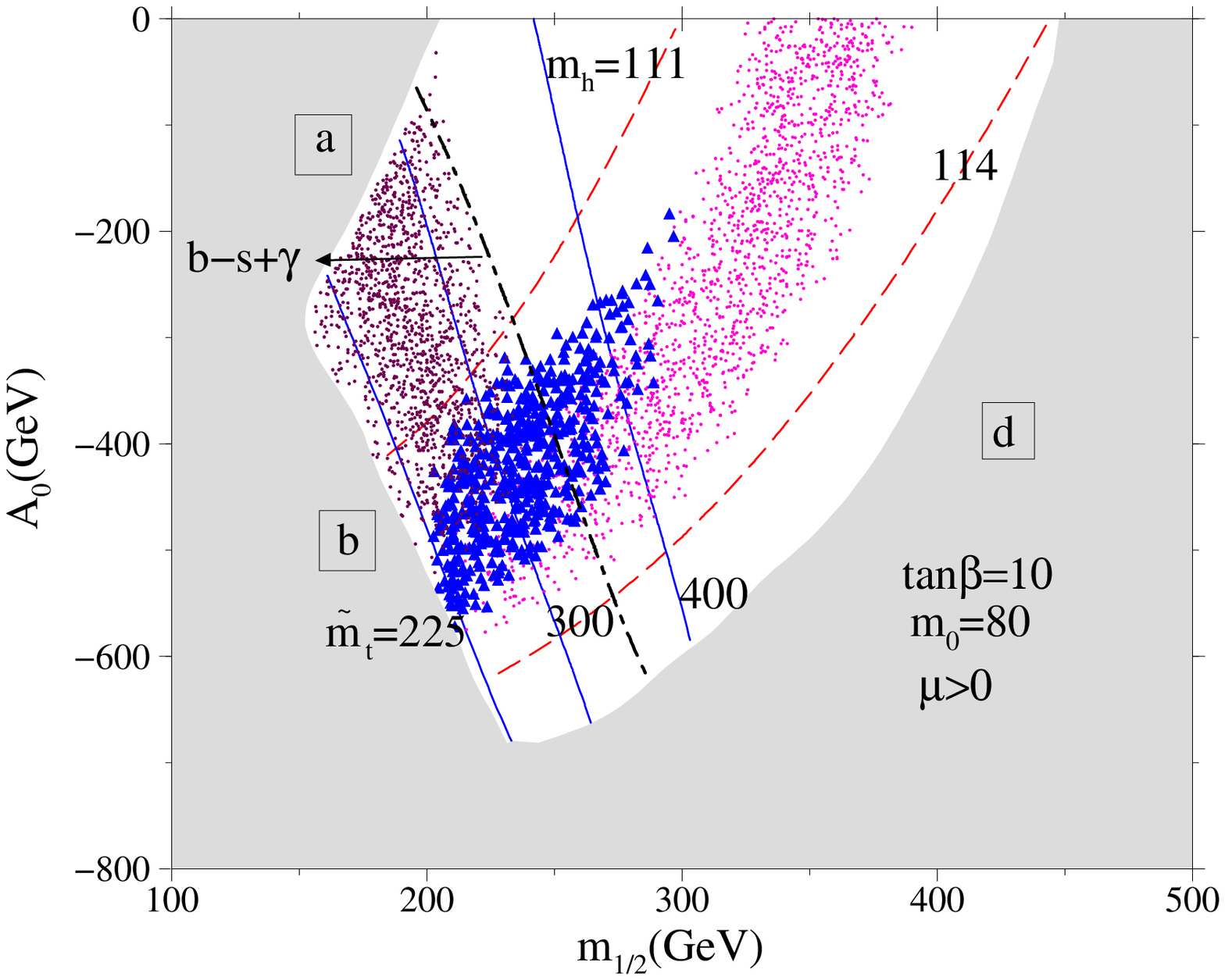}

\hspace*{1.5in}
\mygraph{t10m0170}{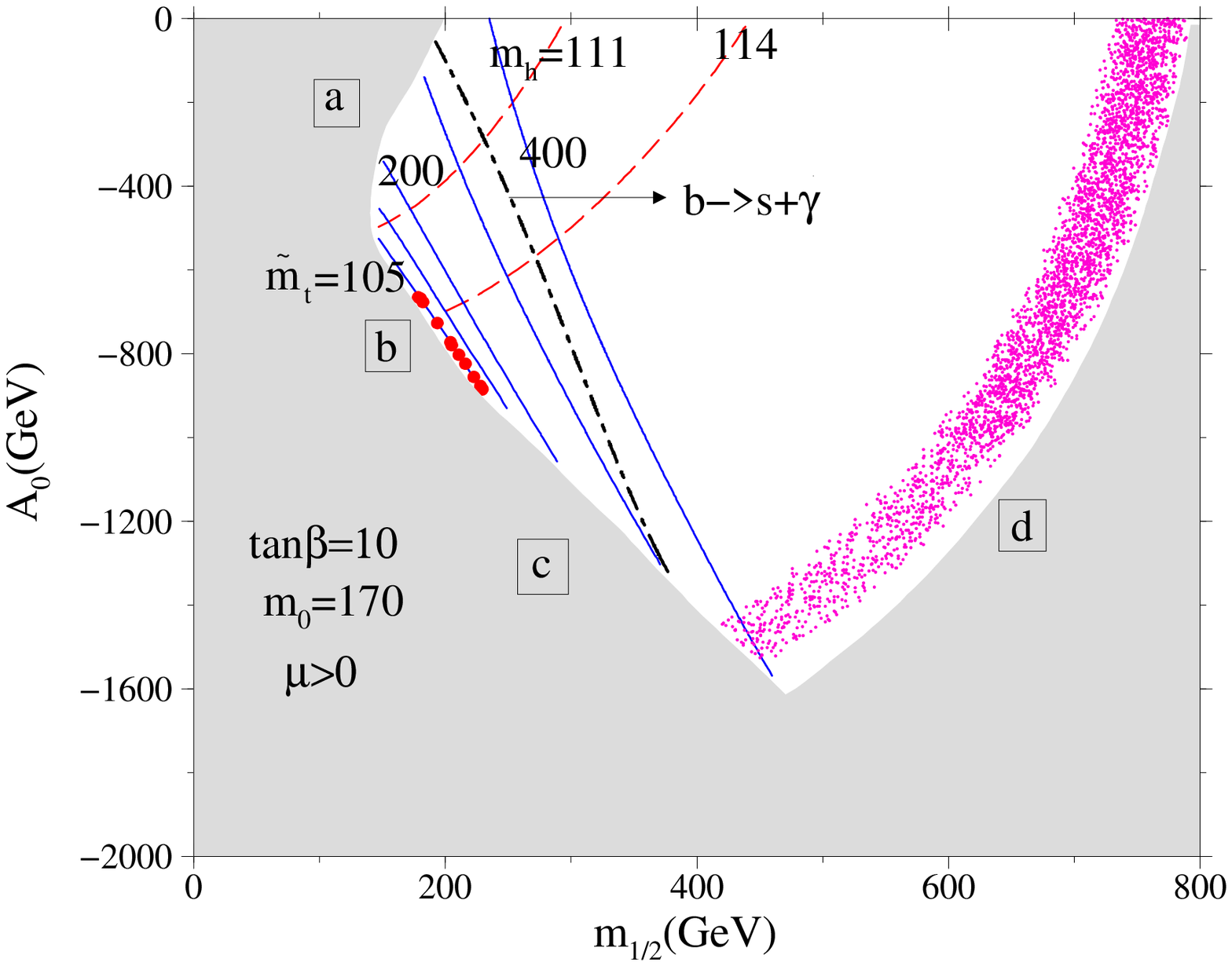}
\caption{
WMAP allowed region in the $m_{1/2}-A_0$ plane for $\tan\beta=10$ 
and $m_0=$~120, 80 and 170~GeV. Lighter stop masses are shown in 
(blue) solid lines. Dashed lines correspond to lighter Higgs masses. 
Dot-dashed lines refer to $b\rightarrow s \gamma$ limits. WMAP
allowed zones: The brown(medium shaded) region near zone~(a) refers
to LSP pair annihilation. 
In the red region 
near zone~(b) LSP pair annihilation and LSP-$\lstop$ coannihilation jointly
produce the relic density. 
The blue(deep shaded) region near zone~(c) refers to 
a mixture of bulk annihilation and LSP-${\tilde \tau}_1$
coannihilation. 
The pink(light shaded) region near zone~(d) corresponds to a dominantly 
LSP-${\tilde \tau}_1$ 
coannihilation. 
Disallowed zones:     
Region (a) is disallowed because either ${\tilde \chi}_1^ \pm$ falls
below the LEP limits or $m_h$ becomes unacceptably small 
($<108$~GeV). 
Region~(b) is ruled out by the lower limits 
on $\mlstop$ or/and $m_{{\tilde \tau}_1}$.
Region~(c) is disfavoured because CCB violating minima occurs at the 
elctroweak 
scale. Region~(d) is discarded as ${\tilde \tau}_1$ becomes LSP. 
}              
         
\label{t10figs}
\end{figure}

\begin{figure}[!ht]
\vspace*{-0.05in}
\mygraph{t5m0120}{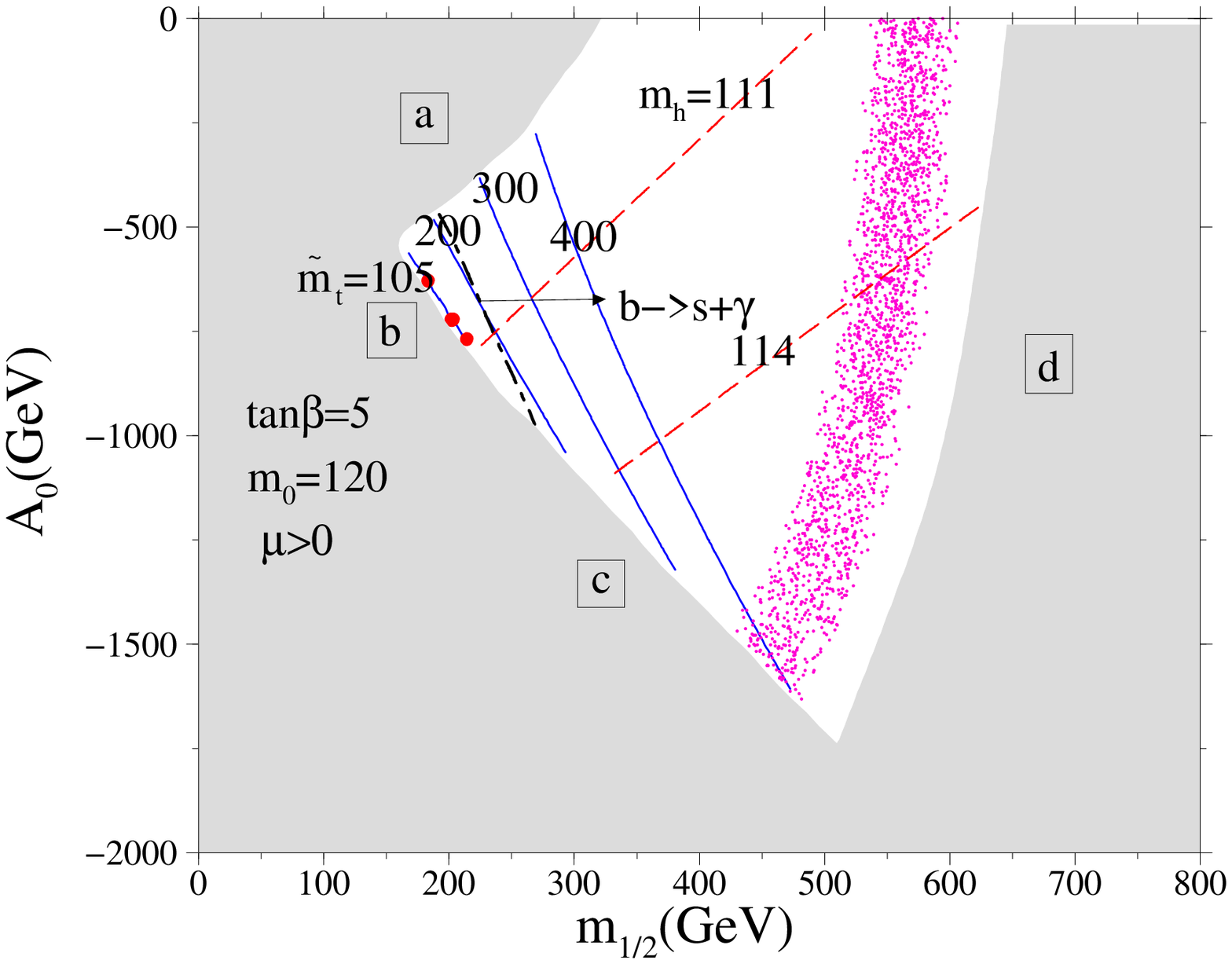}
\hspace*{0.5in}
\mygraph{t5m080}{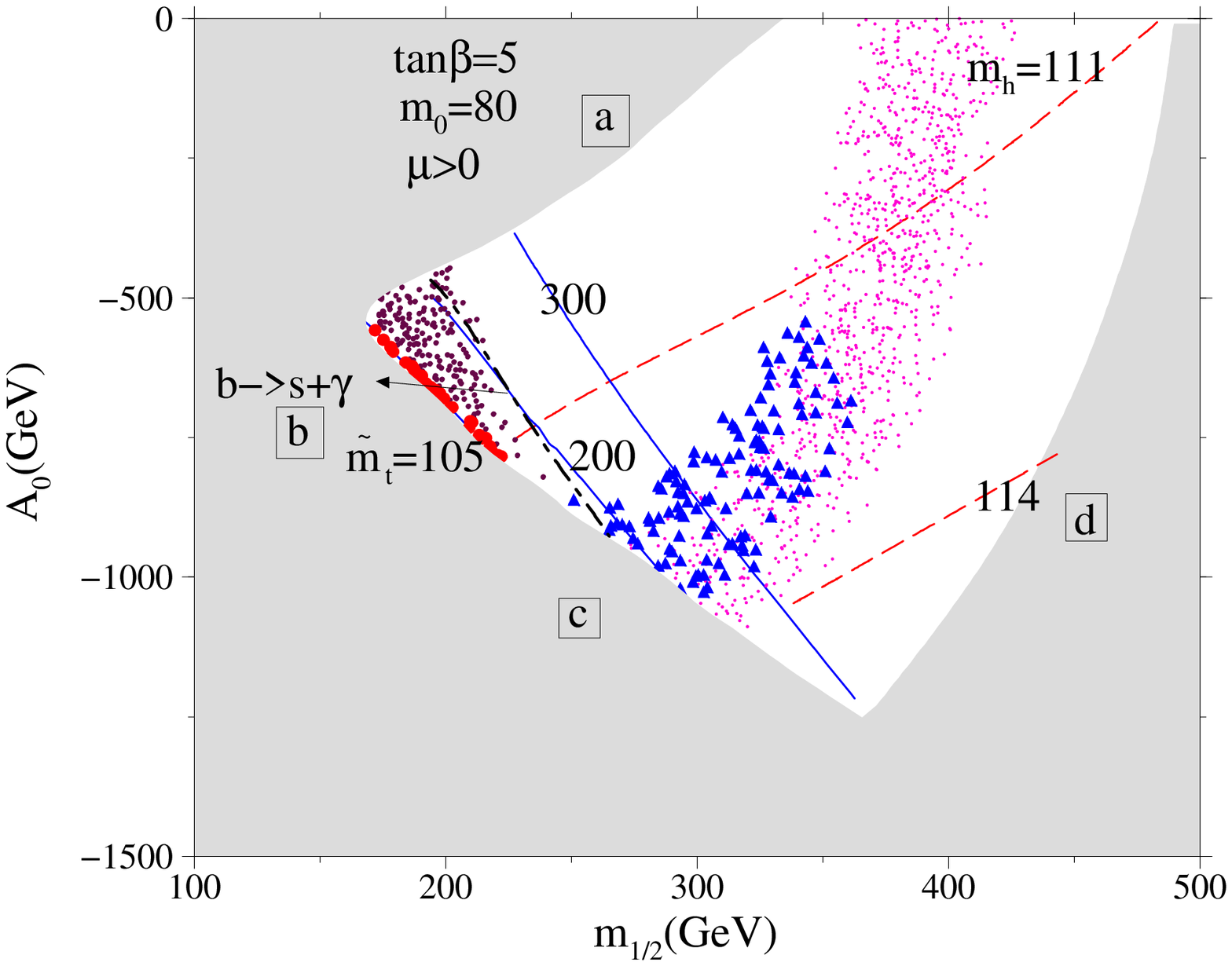}
\hspace*{1.5in}
\mygraph{t5m0170}{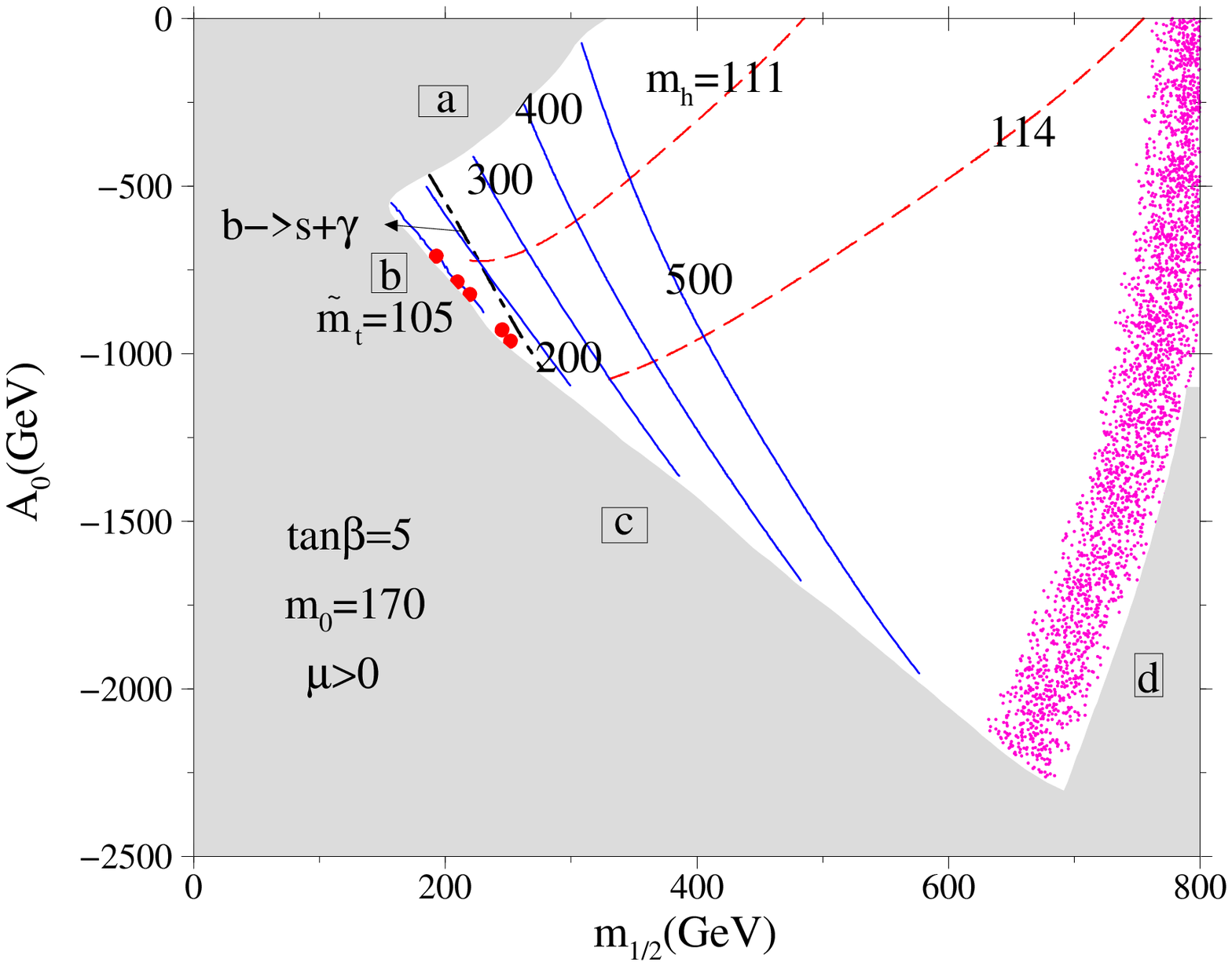}
\caption{
Same as Fig.\ref{t10figs} except for $\tan\beta=5$.
}                       
\label{t5figs}
\end{figure}

\begin{figure}[!ht]
\hspace*{-0.6in}
\centering
\includegraphics[width=0.5\textwidth]{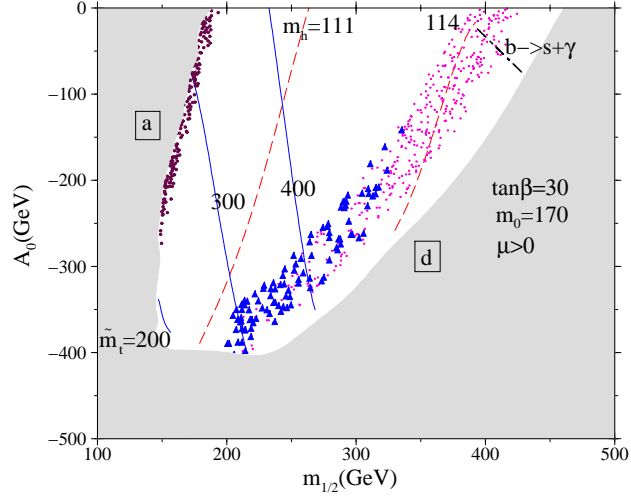}
\caption{Same as Fig.\ref{t10figs} except for $\tan\beta=30$.}
\label{t30fig}
\end{figure}

\begin{figure}[!ht]
\hspace*{-0.6in}
\centering
\includegraphics[width=0.5\textwidth]{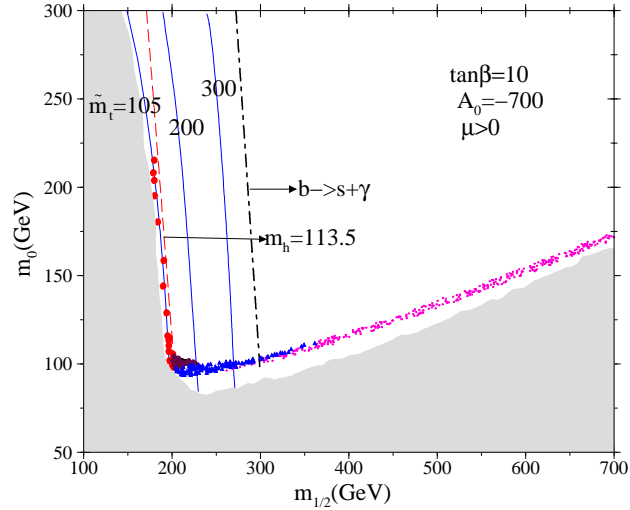}
\caption{Same as Fig.\ref{t10figs} except for being drawn in 
$m_0-m_{1/2}$ plane for $\tan\beta=10$ and $A_0=-700$~GeV.}
\label{t10mhalfm0fig}
\end{figure}

 Additionally, we have randomly varied $m_0$ 
by $5$~GeV around the quoted value of $120$, $80$ and $170$~GeV 
in Fig.\ref{t10figs}, Fig.\ref{t5figs} and Fig.\ref{t30fig} respectively. 
We have introduced this variation since different spectrum generators do 
have small uncertainties in computing the sparticle masses in the mSUGRA framework. 
Similarly we have used a variation of $10$~GeV for $A_0$ in  
Fig.\ref{t10mhalfm0fig} around $A_0=-700$~GeV. 

We have also shown the consequences of the indirect constraint from the measured 
branching ratio of $b \rightarrow s \gamma$ \cite{bsg-recent,bsg-okumura}.  
However as 
mentioned in the section.\ref{intro4} these constraint becomes weaker if some additional 
theoretical assumptions are dispensed with.  We have used the constraint
 

\beq
2.77 \times 10^{-4} < Br (b \rightarrow s \gamma) < 4.33 \times 10^{-4}
\label{bsg-result}
\eeq
\noindent
at three sigma level.


Fig.(\ref{t10m0120}) shows different WMAP allowed zones 
in the $(A_0-m_{1/2})$ plane for $\tan\beta=10$, $m_0=120$~GeV. As 
mentioned 
before we have identified regions corresponding to  
$111~\rm GeV<m_h< 114~{\rm GeV}$ (the uncertain zone) and $m_h> 114~{\rm GeV}$
(the regular zone).    
We find WMAP allowed regions for small $\mhalf$ in 
both the zones. Additionally contours for 
$m_{{\tilde t}_1}$ are also shown. 
The WMAP allowed parameter space has three distinct regions.

\begin{enumerate}
\item In the region marked by the red dots the LSP pair annihilation is the 
dominant mechanism. However, this alone cannot satisfy the relic density 
constraint. Additional contributions come from a significant degree of 
LSP-$\lstop$ coannihilation.   
As already mentioned this coannihilation 
is a direct consequence of a top squark NLSP with mass as low as $\sim$ 105~GeV
due to large mixing effects in the top squark mass matrix for large and 
negative $A_0$.
Such a light $\mlstop$ with $\mlstop - \mlspone \sim$ 30~GeV is 
certainly within the striking range of Run II \cite{demina}. Thus if this 
scenario is indeed the one chosen by nature, the discovery of SUSY at LHC
will be heralded by the discovery of the light top squark at the Tevatron.
 
\item In the region marked by the blue(deep shaded) dots  neutralino pair 
annihilation is the main mechanism for satisfying the observed relic 
density. However there are points, the ones with relatively large $\mhalf$
in particular,  where $\sim 50\%$ of the relic density is due to 
${\tilde \tau}_1$-LSP coannihilation. 

\item There is a third WMAP allowed region for larger $\mhalf$(the pink/light
shaded dots). 
 
Here $\tilde{\tau}_1$-LSP coannihilation dominantes over
LSP annihilation in satisfying the relic density. 
This coannihilation process becomes more and more effective 
as $\mhalf$ increases. 
However as noted before, in a scenario with REWSB, 
a moderately large negative $A_0$ yields large $|\mu|$. This
causes $m_{{\tilde\tau}_1}$ to become smaller for smaller values of 
$m_{1/2}$ in comparison 
to the vanishing $A_0$ case. Thus larger $|A_0|$ may trigger ${\tilde \tau}_1$-LSP coannihilation at smaller values of $m_{1/2}$ which in turn predicts 
lighter squarks and gluinos in the WMAP allowed region.  
\end{enumerate}

The situation changes significantly if we change $m_0$ in either 
direction. Fig.(\ref{t10m080}) illustrates the case of $m_0= 80$~GeV.  
Here $|A_0|$ could not be as large as in the previous case Fig.(\ref{t10m0120}) 
because that would reduce $m_{{\tilde \tau}_1}$ below 
the experimental limit. 
This on the other hand does not allow any appreciable increase of $m_h$ due to
$A_0$. 
Smaller $|A_0|$ does not reduce $\mlstop$ appreciably 
so as to have any LSP-${\tilde t}_1$ coannihilation.  We note in passing that
$m_{\tilde{t_1}} \le $ 200 GeV is disallowed since ${\tilde \tau}_1$ takes the charge of LSP here. 
This of course disfavours any LSP - ${\tilde t}_1$ coannihilation and ${\tilde t}_1$ is
unlikely to be visible at the
Tevatron \cite{demina}.

The pure bulk region (shown in brown) is extended in comparison to the case of 
$m_0=120~{\rm GeV}$ of Fig.(\ref{t10m0120}). However most of the extended parameter space 
corresponding to the bulk region has $m_h<111~{\rm GeV}$ in this case. 
Additionally there is a sizable mixed region (where both LSP pair annihilation and 
LSP-${\tilde\tau}_1$ coannihilation are important) which falls in the uncertain 
$m_h$ zone.

On increasing $m_0$ to 170~GeV (Fig.(\ref{t10m0170})) we find no bulk 
annihilation 
region. Here negative and larger $A_0$ tends to reduce $m_{{\tilde \tau}_1}$, but a larger 
$m_0$ compensates.  Thus ${\tilde t}_1$ may become the NLSP without having 
${\tilde \tau}_1$ below 
${\tilde \chi}_1^0$.  As a consequence of such large $A_0$ one 
finds region with $m_h \ge$114. This scenario 
has a large LSP-${\tilde \tau}_1$ coannihilation region 
(pink/light shaded dots) 
as well as a small 
LSP-${\tilde t}_1$ coannihilation region (red dots). 
However, part of 
the LSP-${\tilde \tau}_1$ coannihilation region is associated with significantly light 
squark/gluino masses compared to the $A_0 =0$ case
and such regions are well within the reach of the early probes at LHC.

Now we mention the status of the b$\ra$ s $ \gamma$ constraint. The constraint
disfavours the LSP-$\tilde{t_1}$ coannihilation region altogether in all the
cases discussed above. But, segments of all three WMAP allowed regions are
consistent with this constraint for $m_0=80~\rm GeV$ whereas a tiny part of
mixed region and broad part of coannihilation region for $m_0=120$~GeV survive
this constraint. On the other hand, for $m_0=170$~GeV the entire LSP-
${\tilde\tau}_1$ coannihilation region is allowed by this constraint. However,
as mentioned in the section.\ref{intro4} the $b\rightarrow s \gamma$ constraint may loose
much of its impact if additional theoretical assumptions like the alignment
of the quark and the squark mass matrices are relaxed. We therefore prefer not to
exclude any region allowed by the WMAP data on the basis of this constraint.

Fig.(\ref{t5figs}) shows the result for $\tan\beta=5$. 
Here, allowed values of $A_0$ in Fig.(\ref{t5figs}) has a range larger than that of Fig.(\ref{t10figs}). The WMAP allowed regions for $m_0=120$~GeV
Fig.(\ref{t5m0120}) consist of only coannihilations of LSP with ${\tilde t}_1$ 
and ${\tilde \tau}_1$ in contrast to the Fig.(\ref{t10m0120}) where a
significant amount of mixed zone also exist. 
The dark matter satisfied parameter space for $m_0=80$~GeV of 
Fig.(\ref{t5m080}) survives only if full allowance is made for the 
theoretical uncertainty in the lighter Higgs mass calculation i.e the WMAP 
allowed region coincides with the uncertain $m_h$ region. 
However, most of the region satisfy the $b\rightarrow s \gamma$ constraint of 
Eq.(\ref{bsg-result}) except the pure bulk region and 
the LSP - ${\tilde t}_1$   coannihilation region.
If this constraint is relaxed light top squarks 
within the striking range of Run II experiments are permitted. However,
search for such light top squarks requires caution. For  low $\tan 
\beta$ the branching ratio (BR) of the 4-body decay of this squark
may be sizable or may even overwhelm the conventional loop decay\cite{boehm}.
Search strategies should be modified to include this usually neglected decay 
mode\cite{me1}.


As an example of a large $\tan\beta$ case Fig.(\ref{t30fig}) shows 
the result for $\tan\beta=30$ and $m_0=170$~GeV. Here the WMAP constraint is 
satisfied via LSP-${\tilde \tau}_1$ coannihilation even if $m_h > $ 114 GeV.
One also obtains a 
mixed region where pair annihilation is the dominant mechanism. However
this region corresponds to the uncertain $m_h$ region i,e 
111~GeV~$<m_h<$~114~GeV. 

We next study the available parameter space 
in the $m_0$ - $\mhalf$ plane for a fixed negative 
value of $A_0=-700$~GeV. As we see from Fig.(\ref{t10mhalfm0fig}) 
we get a pure bulk annihilation region, a mixed region as well as a 
${\tilde \tau}_1$-LSP coannihilation region all satisfying 
the lighter Higgs bound and the WMAP limits. Additionally there is a significant 
region with LSP-${\tilde t}_1$ coannihilation. 

\section{The Novel Collider 
Signals for non-zero values of the trilinear coupling}

In this section
we simulate all possible squark - gluino events using Pythia (version 6.409)
\cite{pythia} at the LHC energy ($\sqrt{s}$=14 TeV). We compute  
the relevant BRs using the program SDECAY \cite{suspect}.
From Fig (\ref{t10m0120}) we have chosen the 
following benchmark points as shown in Table 1.

\noindent 
A)This point corresponds to a mixed region where neutralino pair annihilation
plays the dominant role. B)This point corresponds to the
stau coannihilation region for a large negative value of $A_0$ and C) 
This point corresponds to the lowest value of $\mhalf$ in the LSP- ${\tilde \tau}_1$ coannihilation region for $A_0$ = 0.

\begin{table}[!ht]
\begin{center}\

\begin{tabular}{|c|c|c|c|}
       \hline
mSUGRA &A&B&C\\
parameters & &&\\
\hline
$m_0$ &120.0  &120.0 &120.0\\
\hline
$m_{1/2}$ &300.0  &350.0 &500.0\\
\hline
$A_0$ &-930.0  &-930.0 &0.0\\
\hline
$\tan\beta$ &10.0  &10.0 &10.0\\
\hline
$sgn(\mu)$ &1.0  &1.0 &1.0\\
\hline

\end{tabular}
\end{center}
   \caption{
The three benchmark mSUGRA scenarios: (A):  This parameter point 
with non-vanishing $A_0$ belongs to a mixed 
region characterized by LSP pair annihilation and LSP- ${\tilde \tau}_1$ 
coannihilation, (B): This point with non-vanishing $A_0$ 
belongs to a region where 
stau coannihilation dominates among the annihilation/coannihilation 
processes, (C) This point with vanishing $A_0$ refers 
to the lowest value of $\mhalf$ in the
LSP-${\tilde \tau}_1$ coannihilation region. All
parameters with dimensions of mass are in GeV.
}
\end{table}

The sparticle spectra for the three benchmark points are shown in Table 2.
  Table 3 shows the branching ratios of the dominant decay modes of
gluino and for the squarks of the first two generations.  The branching ratios
 of the third generation of squarks are shown in Table 4.  We have further 
computed the branching ratios of the dominant decay modes of the lighter
chargino and the second lightest neutralino.  The result is shown in
Table 5. 

\begin{table}[!ht]
\begin{center}

\begin{tabular}{|c|c|c|c|}
       \hline
        Squark/ &&  &   \\
   Slepton/Gluino/  &A&B&C    \\
   Gaugino masses  &&&    \\

        \hline
$\wt g $&719.0  &827.0 &1149.0\\
        \hline
$\wt q_L $&670.0  &769.0&1055.0 \\
        \hline

$\wt q_R $&647.0  &739.0 &1015.0\\
        \hline
$\wt t_1 $&286.0  &396.0&807.0 \\
        \hline
$\wt t_2 $&644.0  &728.0 &1016.0\\
        \hline
$\wt b_1 $&558.0  &652.0 &974.0\\
        \hline
$\wt b_2 $&639.0  & 731.0&1009.0\\
        \hline
$\wt l_L $&239.0  &268.0 &355.0\\
        \hline
$\wt \nu_{l_L} $&226.0&255.0 &346.0\\
        \hline
$\wt l_R $&169.0  &182.0 &224.0\\
        \hline
$\wt \tau_1 $&132.0  &148.0&217.0 \\
        \hline
$\wt \nu_{\tau_L}$&218.0  &248.0&345.0 \\
        \hline
$\wt \tau_2 $&241.0  &268.0 &357.0\\
        \hline
$\chonepm $&232.0  & 272.0&386.0\\
        \hline
$\chtwopm $&616.0  & 669.0&649.0\\
        \hline
$\lspone $&121.0  &142.0 &205.0\\
        \hline
$\lsptwo $&232.0 &272.0 &386.0\\
        \hline
$ h $  &117.0 &117.0&115.0\\
        \hline
\end{tabular}
\end{center}
   \caption{ The sparticle spectra in the three scenarios. All masses
are in GeV.}
\end{table}


\begin{table}[!ht]
\begin{center}\
\begin{tabular}{|c|c|c|c|}
       \hline
Decay modes & A &B&C\\
(squark/gluino) & \% & \% & \%\\
\hline
$\wt g \ra \wt q_L q $ &8.8  &10.4&18.0\\
\hline
$\wt g \ra \wt q_R q $ &18.4  &21.2&35.2\\
\hline
 $\wt g \ra \wt b_1 b$&19.4  &18.6&13.8\\
\hline
 $\wt g \ra \wt b_2 b$&5.6  &6.4&9.4\\
\hline
$\wt g \ra \wt t_1 t$ &47.0  &43.0&23.0\\
\hline
$\wt q_L \ra \chonep q$ &66.0  &66.0&64.6\\
\hline
$\wt q_L \ra \lsptwo q$ &32.6  &32.8&32.0\\
\hline
$\wt q_R \ra \lspone q$ &100.0  &100.0&100.0\\
\hline

\end{tabular}
\end{center}
   \caption{The BRs (\%) of the dominant decay modes of the gluinos and
the squarks belonging to the first two generations.}
\end{table}

\begin{table}[!ht]
\begin{center}\
\begin{tabular}{|c|c|c|c|}
       \hline
Decay modes & A &B&C\\
(heavy squark) & \% & \% & \%\\
\hline
$\wt b_1 \ra \lsptwo b$ &12.0  &14.0&19.7\\
\hline
$\wt b_1 \ra \chonem t$ &16.0  &21.5&35.0\\
\hline
$\wt b_1 \ra \lstop W^{-}$ &70.0  &63.4&13.0\\
\hline
$\wt b_1 \ra \ch2m t$ &0.0  &0.0&29.0\\
\hline
$\lstop \ra \chonem b$ &100.0  &49.0&36.0\\
\hline
$\lstop \ra \lspone t$ &0.0  &51.0&29.0\\
\hline
$\lstop \ra \lsptwo t$ &0.0  &0.0&14.0\\
\hline
$\wt b_2 \ra \lspone b$ &28.4  &3.3&24.5\\
\hline
$\wt b_2 \ra \lsptwo b$ &3.9  &4.2&6.7\\
\hline
$\wt b_2 \ra \chonem t$ &5.9  &6.9&12.0\\
\hline
$\wt b_2 \ra \lstop W^{-} $ &61.0  &54.0&13.0\\
\hline
\end{tabular}
\end{center}
   \caption{The BRs (\%) of the dominant decay modes of the squarks belonging to the 
third generation.}
\end{table}

\begin{table}[!ht]
\begin{center}\

\begin{tabular}{|c|c|c|c|}
       \hline
Decay modes & A &B&C\\
(Gauginos) & \% & \% & \%\\
\hline
$\lsptwo \ra \stau_1^{+} \tau^-$ &78.6&46.0&8.4\\
\hline
$\lsptwo \ra \snu_l \nu_l$ &5.2&24.0&33.8\\
\hline
$\lsptwo \ra \snutau \nu_{\tau}$ &15.0&24.0&17.8\\
\hline
$\chonep \ra \lspone W^+$ &2.6&2.6&5.1\\
\hline
$\chonep \ra \snu_l l^+$ &5.4&25.0&36.0\\
\hline
$\chonep \ra \snutau \tau^+$ &16.0&25.0&19.2\\
\hline
$\chonep \ra \stau_1^{+} \nu_{\tau}$ &76.0&44.0&7.9\\
\hline
\end{tabular}
\end{center}
   \caption{The BRs (\%) of the dominant decay modes of the lighter chargino 
and the second lightest neutralino. All sneutrinos decay into the invisible
channel $\nu$ + $\lspone$ in the three cases under study.}
\end{table}


\begin{table}[!ht]
\begin{center}\

\begin{tabular}{|c|c|c|c|}
\hline
& \multicolumn{3}{c|}{$\sigma(\pb)$}\\
\cline{2-4} 
Process&A&B&C\\
\hline
$\wt g \wt g$ &1.41  &0.55 &0.05\\
$\qL \wt g$ & 2.35 &1.08 &0.15\\
$\qR \wt g$ & 2.55 &1.18 &0.16\\
$\qL \qL$ & 0.78 &0.55 &0.13\\
$\qL \qL^*$ &0.18  &0.08 &0.01\\
$\qR \qR$ &0.75  &0.47 &0.11\\
$\qR \qR^*$ &0.22  &0.10 &0.01\\
$\qL \qR$ &0.47  &0.25 &0.05\\
$\qL \qR^* + c.c$ &0.43  &0.21 &0.04\\
$\lstoppair$ &6.12  &1.14 &0.02\\
$\hstoppair$ &0.07  &0.03 &0.004\\
$\lstop \hstopbar + c.c$ &0.001  &0.0002 &1.5$\times 10^{-5}$\\
$\lsbotpair$ &0.17  &0.07 &0.005\\
$\hsbotpair$ & 0.08 &0.03 &0.004\\
\hline
Total &15.58 &5.74 &0.74 \\
\hline

\end{tabular}
\end{center}
   \caption{The production cross sections of all squark-gluino events
studied in this paper.}
\end{table}

Next we present the total squark-gluino production cross-sections for the 
three scenarios in Table 6. These 
lowest order cross sections have been computed
by CalcHEP (version 2.3.7)\cite{calchep}. 
From this table it is obvious that
even for 10 $ \ifb $ of integrated luminosity scenario A) will produce 
a remarkably
large number of squark-gluino events compared to the other two scenarios.
Thus it can be easily tested by the early  LHC runs.
On the otherhand if ${\tilde \chi}_1^0$ - ${\tilde \tau}_1$ coannihilation
happens to be the dominant mechanism for obtaining the present day thermal
abundance and $A_0$ is large negative (scenario B), the size of
the signal may still be more than an order of magnitude  larger than what is 
expected in scenario C)corresponding to $A_0$ = 0.  The relative sizes of 
the total cross sections in the three cases can be qualitatively 
understood from the sparticle spectra as shown
in Table 2.


\begin{table}[!ht]
\begin{center}\

\begin{tabular}{|c|c|c|c|}
       \hline
 &A&B&C\\
\hline
$1 \tau + X_{\tau}$       & 29870   & 11860  & 1340\\
\hline
$1 \mu + X_{\mu}$        & 5274    & 4251   & 1260\\
\hline
$1 e + X_e$          & 5294    & 4232   & 1262\\
\hline
$1 \tau + 0 b + X_{\tau}$ & 10750   & 4581   & 747\\
\hline
$1 \tau + 1 b + X_{\tau}$ & 19      & 13     & 1\\
\hline
$1 \tau + 2 b + X_{\tau}$ & 17099   & 6589   & 507\\
\hline
$1 \tau + 3 b + X_{\tau}$ & 9       & 6      & 0\\
\hline
$1 \tau + 4 b + X_{\tau}$ & 425     & 665    & 80\\
\hline
\end{tabular}
\end{center}
   \caption{The number of semi-inclusive events with one $\tau$, one $e$
and one $\mu$ at the parton level. X stands for all possible final states
excluding the lepton indicated by the subscript. The last five rows indicate
the number of $b$ quarks in the $\tau$ type events. All numbers correspond to 
an integrated luminosity of 10 $\ifb$. }
\end{table}

\begin{table}[!ht]
\begin{center}\

\begin{tabular}{|c|c|c|c|c|}
       \hline
 &A&B&C& $t \bar t$\\
\hline
$1 \tau + X_{\tau}$  & 3113 & 1402 & 239 & 481 \\
\hline
$1 \mu + X_{\mu}$   & 1179 & 1246 & 820 & 1295 \\
\hline
$1 e + X_e$     & 1138 & 1263 & 829 & 1354\\
\hline

\end{tabular}
\end{center}
   \caption{The number of semi-inclusive events with one detected $\tau$ jet, one 
isolated $e$ and one isolated $\mu$ computed by Pythia. The selection criteria and 
the kinematical cuts are given the text. The last column gives the contributions of
the background from $t - \bar{t}$ production.  
}
\end{table}


We next turn to some distinctive  features  of  the 
signals from the three scenarios.
This criteria may be utilized as a confirmation of the underlying SUSY
scenarios that correspond to different mechanisms of neutralino
annihilation for an acceptable relic density.
From the decay tables it is evident that the signals from decaying
sparticles in A) and B)
will contain many more tau-leptons than electrons or muons 
showing thereby a strong departure from ``lepton universality''.  
However, in scenario C) this
universality in the signal will hold approximately. 

The main reason for this ``non-universality'' 
lies in the dominant 2 - body decay modes of
$\chonepm$ and  $\lsptwo$.
It is to be noted that in scenario A) the $\chonepm$  
decays into the lighter stau(with a dominant $\stau_R$ component)
with a large BR. This is due to the fact that the lighter charginos can 
decay 
only into right handed sleptons via two body modes. However this decay is allowed 
only through the small higgsino component of the 
gaugino like chargino. Thus the decay 
into
$\stauone$  dominates. Again since $\snutau$ is lighter than the other 
sneutrinos, $\chonepm$ decays into  $\snutau$ and tau with a sizable BR.  In all three scenarios 
the sneutrino decays invisibly via the mode ${\tilde \nu}_i \rightarrow \nu_i +{\tilde \chi}_1^0$ 
(i=$e,\mu,\tau$).  Thus the second lightest neutralino dominantly decays either
into invisible neutrino-sneutrino pairs or into $\tau$ and 
$\stauone$  with large BRs. As a result there is a significantly large tau excess 
in the semi-inclusive signal $\tau^\pm$+X$_\tau$, (where X$_\tau$ stands all final states without 
a tau) compared to $e^\pm$+X$_e$ or 
$\mu^\pm$+X$_\mu$. 

In the squark 
gluino events $\chonepm$ and  $\lsptwo$ primarily arise from the 
decays of $\sql$ and $\lstop$ decays into $\chonepm$ and $b$ with 100\%
BR. Moreover  since $\lstop$ is much lighter compared to the other
squarks due to the large magnitude of the $A_0$ parameter the 
gluinos 
decay into $\lstop - t$ pairs with a large BR.

Scenario B) has all the above features 
albeit to a lesser extent. It is
particularly important to note that more than 50$\%$ of the $\lstop$s
decays into $t-\lspone$ pairs. The decays of the $t$ tend to restore
lepton universality. Thus the excess of events with $\tau$ leptons
over the ones with $e$ or $\mu$ is reduced to some extent.  

In scenario C) the excess of tau leptons is reduced drastically
as is evident from decay tables.

However, it must be borne in mind that the 
decays of the standard model particles
(mainly $t,Z$ and $W$) present in the signal exhibit lepton universality. 
Moreover, the heavy flavour ($b,c,t$) decays yield more electrons and 
muons than taus. It is therefore essential  to calculate the relative 
abundance  of  different charged leptons  in the signal through
simplified  simulations. 

To begin with
we ignore the difference in the detection efficiencies  of different
leptons. This will be addressed later in the paper.
Again the production and decay of all squark-gluino pairs are 
generated by Pythia. 
In this simplified parton level analysis all SM  particles other than the $W,Z$  and $t$ are treated as stable. 
All unstable  sparticles 
are allowed to decay. Hadronization, fragmentation and jet formation are
switched off. We also impose the cuts
$p_T>30.0$ GeV and $|\eta_l|<2.5$ on $e, \mu$ and $\tau$. This is 
to have a feeling for the numbers when $e$'s and the $\mu$'s in the final 
state from $b$ and $c$ decays are minimized (see the improved analysis
presented below). 
The results are shown in Table 7. In this table the first row corresponds to the 
number 
of semi-inclusive events with only one $\tau$. The second and third rows 
give the
corresponding numbers for final states with one $e$ and one $\mu$ respectively.

From Table 7 it is obvious that in scenario A) the number of events with
only one $\tau$ is much larger than the corresponding events with $e$ or 
$\mu$. The excess of events involving only one $\tau$ lepton is also
seen in scenario B). The scenario C) exhibits lepton universality.

Since the gluinos dominantly decay into $b$ or $t$ flavoured squarks in 
scenarios A) and B) we expect a sizable number of $b$ quarks in the final 
state. This is also illustrated  in Table 7 (see the last five rows).

Thus in principle $b$-tagging
may be effectively used to suppress the backgrounds except the ones from
$t-\bar{t}$ production.

To demonstrate  that the above tau excess survives the 
typical kinematical cuts designed for  SUSY search experiments at the LHC
we proceed as follows. 
Since 
a detailed background simulation is beyond the scope of this paper we 
apply a slightly modified version of the 
generic cuts introduced by the ATLAS collaboration to
eliminate the SM backgrounds in their study of  
inclusive SUSY signals \cite{atlas}. We shall, however, study the 
response of the $t - \bar t$ events, likely to be the 
dominant SM background,  
to these cuts after our signal analysis.

Again all squark-gluino events are generated by Pythia. 
Initial and final state radiation, decay, hadronization, fragmentation
and jet formation are implemented following the standard procedures in 
Pythia.  We impose the following selection and background rejection criterion:

\begin{enumerate}
\item Only jets having transverse energy $ E_T^{j} >$ 30 GeV, 
pseudo rapidity $|\eta_j|<4.5$ and the  
jet-jet isolation parameter $\Delta R(j_1,j_2)>$ 0.5 are selected 
from the toy calorimeter of Pythia.

\item The jets arising from hadronic tau decays are selected  
as follows. Firstly  
$\tau$'s  having transverse momentum $ p_T >$ 20 GeV and 
pseudo rapidity $|\eta_{\tau}|<3$
are identified. These $\tau$s are matched with the jets by 
setting
the following criterion $\Delta R(\tau,j)<$ 0.4 and 
$ E_T^{j} / E_T^{\tau} >$ 0.8, where $ E_T^{j}$ is the transverse energy 
of 
jets and $E_T^{\tau}$ is the same for the $\tau$ being matched. 
The possibility of detecting the
matched $\tau$s is then computed using the
detection efficiencies  in \cite{cms}(see Fig 12.9, p 444  in section 
12.1.2.2).

\item We also require that an isolated e or
$\mu$  in an event have $p_T >$ 30 GeV and 
the lepton-jet isolation parameter 
$\Delta R(l,j)>$ 0.5, where $l$ stands for either  $e$ or $\mu$.
The detection efficiencies of these leptons are assumed to be 100\%.

\item We reject events without  at least two jets having 
transverse momentum $P_T >$ 150 GeV.

\item Events are rejected if  missing transverse energy $\etslash <$  
200 GeV.

\item Only events with jets having $S_T >$ 0.2, where $S_T$ is a standard 
function of the eigen values of the sphericity tensor,   
are accepted.

\end{enumerate}


The strong $p_T$ and isolation criteria on $e$ and $\mu$ 
enable us to exclude the most of leptons 
from semileptonic $b$ and $c$ decays 
from the list of isolated leptons in an event. We have checked that
a suitable cut on $M_{eff}$, 
where $M_{eff}= |\met| + \Sigma_{i=1}^2|p_T^{l_i}| + \Sigma_{i=1}^4|p_T^{j_i}|$
($l = e,\mu$) does not improve the ratio $S/\sqrt{B}$ significantly, where $S$ 
corresponds to number of signal events and $B$ represents the number of background events.

We present the number of semi-inclusive
events with only one detected $\tau$ jet, with only one isolated 
$\mu$ or $e$ in Table 8. 
The number of events corresponds to an 
integrated luminosity of 10 $\ifb$. It is clear from the table
that inspite of non-ideal $\tau$ detection efficiencies
there is indeed an excess of events with tau leptons over that with electrons 
and muons in 
scenario A). In scenario B) the number of events with tau leptons is 
reduced as expected compared to
A) but is still significantly  larger than the corresponding number 
in C).

We note in passing that there are several proposals for detecting the 
taus \cite{cms}.
The combined efficiency may be larger than the efficiency 
obtained from tracker isolation alone which  we have used.
However, the combined efficiency is process dependent \cite{cms}and in 
this
simplified analysis we have not employed it. Also following the CMS  
guidelines we have not considered the possibility of detecting tau-jets 
with
$p_T<$ 30 GeV. However, we have noted that a large number of tau-jets 
in the final state have
$p_T$ in this region. If future works establish the techniques  of detecting
these relatively low $p_T$ tau jets, the excess of final states involving  
$\tau$s will be even more dramatic.  

We next study the dominant background from $t -\bar t$ events using the 
selection criteria and kinematical cuts  given above.
The leading order cross-section
for this process computed by CalcHEP is 400$\pb$.

 The results are 
given in the last column of Table 8. The $S/\sqrt{B}$ ratios
in the three scenarios  indicate that
statistically significant signals are expected in all cases. Since 
$b$-tagging is not particularly helpful in rejecting this particular 
background we have not incorporated it here. However in a more complete
background analysis $b$-tagging may indeed be a useful tool as our parton 
level analysis indicates.   

As noted in the introduction a consequence of large negative $A_0$ is the 
possibility of a light top squark which can be as light as the 105 GeV
(Fig.\ref{t10figs} - \ref{t10mhalfm0fig}). 
For such a light $\lstop$ its coannihilation with the LSP 
may be one of the important mechanisms for having the observed DM in the universe.
Moreover, the LSP mass in this region of the parameter space is about 25-30 GeV
lesser than $m_{{\tilde t}_1}$.
Hence, the top squark NLSP 
will decay via the loop mediated mode $c + \lspone$ with 100
\% BR.
The prospect of light stop squark search in this channel 
at Tevatron Run II  was studied in \cite{demina}. If this scenario 
for thermal abundance of 
dark matter in the universe is realized in nature SUSY
search at LHC will certainly be heralded by the discovery of $\lstop$
at the Tevatron. Even in scenario B) the $\lstop$ will be considerably
lighter compared to scenario C) (see Table 2) although its mass will 
still be outside the kinematic reach of the Tevatron. However this squark will 
have a sizable production cross section at the LHC. Hence its discovery 
may be an additional confirmation of a mechanism of producing 
an acceptable DM relic density based on
${\tilde \chi}_1^0$ - ${\tilde t}_1$ coannihilation with large negative $A_0$.  

An additional feature of WMAP allowed regions of the parameter space with low $m_0$ is 
the existence of sneutrinos decaying invisibly. This happens in all three scenarios A, B 
and C. Moreover, the $\lsptwo$ decays into sneutrino neutrino pairs with sizable BR. Thus 
in addition to the LSP there are two other carriers of missing energy.  An important 
consequence of these scenarios is the spectacular enhancement of the signal 
$e^+e^-$ 
$\rightarrow \gamma$ + missing energy over the SM background from $e^+e^-$ $\rightarrow 
\nu_i \bar {\nu_i} + \gamma$ (i=$e,\mu, \tau$). The signal comes from three channels 
$e^+e^-$$\rightarrow {\tilde \chi}_1^0 {\tilde \chi}_1^0 \gamma$,${\tilde \chi}_2^0 
{\tilde \chi}_2^0 \gamma$, ${\tilde \nu}{{\tilde \nu}^\star} \gamma$. The signal from the 
first SUSY process was computed in \cite{pandita} with the approximation that ${\tilde 
\chi}_1^0$ 
is a pure photino.  All three cross-sections including the effect of 
neutralino mixing 
were computed in \cite{sreerup}.  The results of \cite{sreerup} were compared with that of 
CalcHEP in 
\cite{asesh}. The observation of the signal at the ILC may provide additional confirmation 
of 
SUSY dark matter if produced in the low $m_0,m_{1/2}$ region of the mSUGRA model.



\section{Conclusions}

In this work we observe that parts of the low $m_0$ - $m_{1/2}$ region of the mSUGRA model 
with moderate or large negative values of the common trilinear coupling 
$A_0$ at $M_G$ are consistent with the current direct bounds on sparticle 
masses from LEP as well as the WMAP bound on the dark matter relic density 
(Fig.\ref{t10figs} - \ref{t10mhalfm0fig}).
 In particular the LEP bound $m_h >$114 GeV can be easily 
satisfied for relatively small $m_0$ and $m_{1/2}$ if $A_0$ is large 
and negative 
(Fig.\ref{t10figs} - \ref{t10mhalfm0fig}). This possibility is excluded by the 
adhoc choice $A_0$ =0 made in many current analyses.

The relic density is produced by several processes including
LSP - pair annihilation, LSP - ${\tilde t}_1$ or LSP - ${\tilde
\tau}_1$ coannihilation or by suitable combinations  of these processes.

This parameter space corresponding to relatively light squarks and light 
gluinos 
is important since it will be extensively scanned at the early runs of the 
LHC.  A large excess of semi-inclusive events of the type 
$\tau^\pm$+X$_\tau$ (see Tables 7 and 8) over $e^\pm$+X$_e$ or 
$\mu^\pm$+X$_\mu$ 
events is a hallmark of scenarios with large negative 
$A_0$ in mSUGRA. Our 
simulations with three bench-mark scenarios (see Table 1) consistent with 
direct lower bounds on sparticle masses and WMAP data establish this for a 
modest integrated luminosity of 10 $\ifb$.

A natural consequence of large negative $A_0$ is a light ${\tilde t}_1$.  If that 
be the case then ${\tilde t}_1$ - ${\tilde \chi}_1^0$ coannihilation may 
be an important mechanism for the present day dark matter density.  In this scenario 
the discovery of SUSY at the LHC is likely to be preceded by the 
discovery of ${\tilde t}_1$ at the Tevatron.

Several groups have also imposed indirect constraints on the allowed 
parameter space \cite{dmsugra}. However, it is well known that each of these 
indirect constraints employ additional theoretical assumptions which are 
not fool proof.  For example the requirement that there be no CCB minima 
of the scalar potential deeper than the EWSB vacuum \cite{casasCCB} becomes 
redundant if the latter is assumed to be a false vacuum with a life time 
larger than the age of the universe \cite{false}.  Similarly the constraint from 
the measured branching ratio of the decay $b \rightarrow s \gamma$ derived 
under the purely theoretical assumption that quark and squark mass 
matrices are perfectly aligned can also be evaded if this assumption is 
relaxed ( see,e.g., Djouadi et al in \cite{dmsugra}and references therein).  
Although we have shown the 
consequences of some of these 
indirect bounds in our figures it may not be prudent to discard any parameter 
space 
if these bounds are violated.

{\bf Acknowledgment}: 
AD acknowledges financial support from the Department of Science and Technology,
Government of India under the project  No (SR/S2/HEP-18/2003).
SP and DD
would like to thank the Council of Scientific and
Industrial Research, Govt. of India for financial support.


\end{document}
